\documentclass[format=acmsmall, review=false, screen=true]{acmart}

\setcopyright{rightsretained}

\usepackage{graphicx}
\usepackage{balance}
\usepackage{xcolor}
\usepackage{url}
\usepackage{amssymb}
\usepackage{amsmath}
\usepackage{multirow}
\usepackage{booktabs}
\usepackage{makecell}
\usepackage{natbib}
\usepackage[caption=false]{subfig}
\usepackage{wrapfig}

\usepackage[noend]{algpseudocode}
\usepackage[linesnumbered,vlined]{algorithm2e}
\SetKwFor{For}{{\texttt{for}}}{}{}
\SetKwFor{While}{{\texttt{while}}}{}{}
\SetKwIF{If}{ElseIf}{Else}{{\texttt{if}}}{}{\texttt{else if}}{{\texttt{else}}}{}%
\newcommand{\code}[1]{\texttt{\textbf{#1}}}
\newcommand{\Ret}{\code{return}}

\usepackage{xcolor, soul}
\usepackage{siunitx}
\usepackage{enumitem}
\usepackage{hyperref}
                                
\definecolor{DarkGray}{gray}{.45}

\newcommand{\parag}[1]{\return\noindent\textbf{\textsf{#1.}}\,}
\newcommand{\return}{\vspace{0.2cm}}
\DeclareMathOperator{\msp}{\mbox{ }}
\newcommand{\Oh}{O}
\newcommand{\seq}{\mathcal{S}}

\newcommand{\nonl}{\renewcommand{\nl}{\let\nl\oldnl}}

\newcommand{\pr}{\mathbb{P}}

\newcommand{\google}{\textsf{GoogleV2}}
\newcommand{\yahoo}{\textsf{YahooV2}}
\newcommand{\euro}{\textsf{Europarl}}
\newcommand{\BW}{\textsf{1BillionWord}}
\newcommand{\WP}{\textsf{Wikipedia17}}

\newcommand{\CW}{\textsf{ClueWeb09}}

\newcommand{\sel}{\textsf{select}}
\newcommand{\lk}{\textsf{lookup}}

\newcommand{\bpg}{\textsf{bytes/gram}}
\newcommand{\mpq}{\textsf{$\mu$sec/query}}

\newcommand{\EF}{\textsf{EF}}    % Elias-Fano
\newcommand{\PEF}{\textsf{PEF}}  % partitioned Elias-fano
\newcommand{\MPH}{\textsf{MPH}}
\newcommand{\spaceopt}{\textsf{PEF-RTrie}}
\newcommand{\timeopt}{\textsf{PEF-Trie}}

\newcommand{\ken}{{\sf KenLM}}
\newcommand{\berk}{{\sf BerkeleyLM}}
\newcommand{\rand}{{\sf RandLM}}
\newcommand{\mari}{{\sf Marisa}}
\newcommand{\expg}{{\sf Expgram}}

\newcommand{\os}{\textup{\textsf{1-Sort}}}
\newcommand{\ts}{\textup{\textsf{3-Sort}}}

\makeatletter
\def\@subsubsecfont{\sffamily\bfseries}
\renewcommand\subsubsection{\@startsection{subsubsection}{3}{0pt}%
  {-.5\baselineskip \@plus -2\p@ \@minus -.2\p@}%
  {-3.5\p@}%
  {\@subsubsecfont\@adddotafter}}
\makeatother

\acmJournal{TOIS}
\acmYear{2019}
\copyrightyear{2019}
\setcopyright{acmlicensed}
\acmDOI{https://doi.org/10.1145/3302913}

\begin{document}

\title{Handling Massive $N$-Gram Datasets Efficiently}

\author{Giulio Ermanno Pibiri}
\author{Rossano Venturini}

\affiliation{
  \institution{\\University of Pisa and ISTI-CNR}
  \city{Pisa}
  \country{Italy}
}

\email{giulio.pibiri@di.unipi.it}
\email{rossano.venturini@unipi.it}

\begin{abstract}

\medskip

Two fundamental problems concern the handling of large $n$-gram language models:
\emph{indexing}, that is compressing the $n$-grams and associated satellite values without compromising their retrieval speed, and
\emph{estimation}, that is computing the probability distribution of the $n$-grams extracted from a large textual source.

Performing these two tasks \emph{efficiently} is vital for several applications in the fields of Information Retrieval, Natural Language Processing and Machine Learning, such as auto-completion in search engines and machine translation.

%Because of the stringent efficiency requirements, dealing with billions of $N$-grams poses the challenge of introducing a compressed representation that preserves the query processing speed.
%In this paper we study the problem of reducing the space required by the
%representation of such datasets, maintaining the capability of looking up for
%a given $N$-gram within micro seconds.

\return
Regarding the problem of indexing, we describe compressed, exact and lossless data structures that achieve, at the same time, high space reductions and no time degradation with respect to the state-of-the-art solutions and related software packages.
In particular, we present a compressed trie data structure in which each word of an $n$-gram following a \emph{context} of fixed length $k$, i.e., its preceding $k$ words, is encoded as an integer whose value is proportional to the number of words that follow such context.
Since the number of words following a given context is typically very
small in natural languages, we lower the space of representation to compression levels that were never achieved before, allowing the indexing of billions of strings.
Despite the significant savings in space, our technique introduces a negligible penalty at query time.

Specifically, the most space-efficient competitors in the literature, that are both quantized and lossy, do not take less than our trie data structure and are up to 5 times slower. Conversely, our trie is as fast as the fastest competitor, but also retains an advantage of up to $65\%$ in absolute space.

\return
Regarding the problem of estimation, we present a novel algorithm for estimating \emph{modified Kneser-Ney} language models that have emerged as the de-facto choice for language modeling in both academia and industry, thanks to their relatively low perplexity performance.
Estimating such models from large textual sources poses the challenge of devising algorithms that make a parsimonious use of the disk.

The state-of-the-art algorithm uses three sorting steps in external memory: we show an improved construction that requires only one sorting step by exploiting the properties of the extracted $n$-gram strings.
With an extensive experimental analysis performed on billions of $n$-grams, we show an average improvement of $4.5$ times on the total running time of the previous approach.

\bigskip

\end{abstract}




\maketitle

\renewcommand{\shortauthors}{G. E. Pibiri and R. Venturini}

\newpage

\section{Introduction}\label{sec:intro}

An \emph{$n$-gram} is a sequence of $n$ tokens, where $n$ ranges from $1$ to $N$ that is a small constant, e.g., $N = 5$.
A token can be either a single character or a word, the latter intended as a sequence of characters delimited by a special symbol (e.g., a whitespace character).
The use of $n$-grams is wide and vital for many tasks in Information Retrieval, Natural Language Processing and Machine Learning, such as: auto-completion in search engines~\cite{BYK11,mitra14,mitra15}, spelling correction~\cite{Kukich92}, similarity search~\cite{kondrak2005n}, identification of text reuse and plagiarism~\cite{JC08,HMC11}, automatic speech recognition~\cite{NLP-book} and machine translation~\cite{Heafield11,PK11}, to mention some of the most notable.

For example, query auto-completion is one of the key features that (virtually) any modern search engine offers to help users formulate their queries. The objective is to predict the query by saving keystrokes: this is implemented by reporting the top-$k$ most frequently-searched $n$-grams that follow the keywords typed by the user~\cite{BYK11,mitra14,mitra15}.
The identification of such patterns is possible by traversing a data structure that stores the $n$-grams seen during previous user searches.
Given the number of users served by large-scale search engines and the high query rates, it is of utmost importance that such data structure traversals are carried out in a handful of microseconds~\cite{NLP-book,BYK11,CMS09,mitra14,mitra15}.
Another noticeable example is spelling correction in text editors and web search. In their basic formulation, $n$-gram spelling correction techniques work by looking up every $n$-gram in the input string in a pre-built data structure in order to assess their existence or return a statistic, e.g., a frequency count, to guide the correction~\cite{Kukich92}.
If the $n$-gram is not found in the data structure it is marked as a misspelled pattern: in such case, the correction happens by suggesting the most frequent word that follows the pattern with the longest matching history~\cite{NLP-book,Kukich92,CMS09}.

At the core of such example applications lies an \emph{efficient} data structure mapping the $n$-grams to their associated satellite values, e.g., the frequency counts representing the number of occurrences of the $n$-grams in some domain of interest or probability/backoff weights for word-predicting computations~\cite{Heafield11,PK11}.
The efficiency of the data structure should be both in time \emph{and} space, because modern string search and machine translation systems make many queries over databases containing several billion $n$-grams that often do not fit in internal memory~\cite{NLP-book,CMS09}.
To reduce the memory-access rate and, \emph{hence}, speed up the execution of the retrieval algorithms, the design of an efficient \emph{compressed} representation of the data structure appears as mandatory.
While several solutions have been proposed for the indexing and retrieval of $n$-grams, either based on \emph{tries}~\cite{trie} or \emph{hashing}~\cite{hashing}, their practicality is actually limited because of some important inefficiencies that we discuss below.

Context information, such as the fact that \emph{relatively few} words may follow a given context, is not currently exploited to achieve better compression.
When query processing speed is the main concern, space efficiency is almost completely neglected by not compressing the data structure using sophisticated encoding techniques~\cite{Heafield11}.
In fact, space reductions are usually achieved by either: lossy quantization of satellite values, or by randomized approaches with false positives allowed~\cite{TO07}.
The most space-efficient and lossless proposals in the literature still employ binary search over the compressed representation to search for an $n$-gram: this results in a severe limitation during query processing because of the lack of a compression strategy with a fast random access operation~\cite{PK11}.
To support random access, current methods leverage on \emph{block-wise compression} with expensive decompression of a block every time an element of the block has to be accessed.
Finally, storing $n$-grams in hash tables with linear probing results in a prohibitive space usage since the tables are allocated with a significant extra empty space (e.g., 30-50$\%$) to allow fast random access~\cite{Heafield11,PK11}.

Since a solution that is compact, fast and lossless at the same time is still missing, the first objective of this paper is the one of addressing the aforementioned inefficiencies by introducing compressed data structures that, despite their small memory footprint, support efficient random access to the satellite $n$-gram values. We refer to such problem as the one of \emph{indexing} $n$-gram datasets. 

The other related problem that we study in this paper is the one of computing the probability distribution of the $n$-grams extracted from large textual collections. We refer to this second problem as the one of \emph{estimation}.
In other words, we would like to create an efficient, compressed, index that maps the $n$-grams extracted from the collection to their probabilities.
Clearly, the way such probabilities are computed depends on the chosen model.
The problem of estimation has received a lot of attention throughout the years: not surprisingly, 
several models have been proposed in the literature, such as Laplace, Good-Turing, Katz, Jelinek-Mercer, Witten-Bell and Kneser-Ney (see~\cite{CG96,CG99} and references therein for a complete description and comparison).

Among the many, \emph{Kneser-Ney} language models~\cite{KN95} and, in particular, their \emph{modified} version introduced by~\citet*{CG99}, have gained popularity thanks to their relatively low-perplexity performance. This makes modified Kneser-Ney the de-facto choice for language model toolkits.
In fact, all the following software libraries, widely used in both academia and industry (e.g., Google~\cite{Brantz07,CS13} and Facebook~\cite{CGA16}), support modified Kneser-Ney smoothing:
$\ken$~\cite{Heafield11,HPCK13}, $\berk$~\cite{PK11}, $\rand$~\cite{TO07}, $\expg$~\cite{Watanabe09}, \textsf{MSRLM}~\cite{MSRLM}, \textsf{SRILM}~\cite{SRILM}, \textsf{IRSTLM}~\cite{IRSTLM} and the recent approach based on suffix trees by~\citet{shareghi2015compact,shareghi2016fast}.
For such reasons, Kneser-Ney is the model we consider in this work and that we describe in the background Section~\ref{sec:background}.

The current limitation of the mentioned software libraries is that estimation of such models occurs in internal memory and, as a result, these are not able to scale to the dimensions we consider in this work.
An exception is represented by the work of~\citet{HPCK13} (and included as part of the library $\ken$) that contributed an estimation algorithm involving three steps of sorting in external memory.
Their solution embodies the current state-of-art solution to the problem:  the algorithm takes, on average, as low as $20\%$ of the CPU and $10\%$ of the RAM of the other toolkits~\cite{HPCK13}.
Therefore, our work aims at improving upon the I/O efficiency of this approach.

%\subsection{Our contributions}
\parag{Our contributions}
We list here the contributions of this paper.
\begin{enumerate}[leftmargin=*]
\item We introduce a compressed trie data structure in which each level of the trie is modeled as a monotone integer sequence that we encode with \emph{Elias-Fano}~\cite{Elias74,Fano71} as to efficiently support random access operations and successor queries over the compressed sequence.

As a side contribution, we adopt a hashing approach that leverages on \emph{minimal perfect hash} in order to use tables of size \emph{equal} to the number of stored $n$-grams and spend one random access to retrieve the corresponding $n$-gram satellite value.

\item We describe a technique for lowering the space usage of the trie data structure, by reducing the magnitude of the integers that form its levels.
Our technique is based on the observation that \emph{few} distinct words follow a predefined context, in \emph{any} natural language.
In particular, each word following a context of fixed length $k$, i.e., its preceding $k$ words, is encoded as an integer whose value is proportional to the number of words that follow such context.

\item We present an extensive experimental analysis to demonstrate that our technique offers a significantly better compression with respect to the plain Elias-Fano trie, while only introducing a slight penalty at query processing time.
Specifically, the most space-efficient proposals in the literature, that are both quantized and lossy, do not take less space than our trie data structure and are up to $5\times$ slower.
Conversely, our trie data structure is as fast as the fastest competitor, but also retains an advantage of up to $65\%$ in absolute space.

\item We design a faster estimation algorithm that requires only one step of sorting in external memory, as opposed to the state-of-the-art approach~\cite{HPCK13} that requires three steps of sorting.
The result is achieved by the careful exploitation of the properties of the extracted $n$-gram strings.
Thanks to such properties, we show how it is possible to perform the whole estimation on the \emph{context}-sorted strings and, yet, be able to efficiently lay out the reverse trie data structure, indexing such strings in \emph{suffix} order.
We show that saving two steps of sorting in external memory yields a solution that is $2.87\times$ faster on average than the state-of-the-art approach.

\item We introduce many optimizations to further enhance the running time of our algorithm, such as: asynchronous CPU and I/O threads, parallel least-significant-digit (LSD) radix sort, block-wise compression and multi-threading.
With an extensive experimental analysis conducted over billions of strings, we study the behaviour of our algorithm at each step of estimation; quantify the impact of the introduced optimizations and consider the comparison against the state-of-the-art.
The devised optimizations further improve the running time by $1.6\times$ on average, making our algorithm $4.5\times$ faster than the state-of-the-art solution.
\end{enumerate}

\parag{Paper organization}
Although the two problems we address in this paper, i.e., indexing and estimation, are strictly correlated, we address them one after the other (Section~\ref{sec:indexing} and~\ref{sec:estimation} respectively) in order to introduce the whole material in an incremental way.
In particular, we show the experimental evaluation right after the description of our techniques for each problem, rather than deferring it to the end of the paper.
We believe this form is the most suitable to document the achieved results.
In our intention, each section is an independent unit of exposition.
In the light of these considerations, the paper is structured as follows.

Section~\ref{sec:background} fixes the notation used in the paper and introduces the relevant background, such as the the Kneser-Net smoothing technique.
Section~\ref{sec:related} discusses the state-of-the-art solutions for the two problems we tackle in the paper.
Section~\ref{sec:indexing} treats the problem of indexing.
In particular, Section~\ref{subsec:indexing:ef-trie} describes our compressed trie data structure, whereas Section~\ref{subsec:indexing:hashing} describes the hash-based index. The efficiency of these data structures is validated in Section~\ref{subsec:indexing:exp} with a rich set of experiments.
Section~\ref{sec:estimation} treats the problem of estimation.
In particular, we present our improved estimation algorithm in Section~\ref{subsec:estimation:1-sort} and validate its performance in Section~\ref{subsec:estimation:exp} by also introducing many optimizations.
We conclude the paper in Section~\ref{sec:conclusions}.

\section{Background and Notation}\label{sec:background}

A \emph{language model} (LM) is a probability distribution $\pr(\mathcal{S})$ that describes how often a string $w_1^n = w_1 \cdots w_n$ drawn from the set $\mathcal{S}$ appears in some domain of interest.
The primary goal of a language model is to compute the probability of the word $w_n$ given its preceding history of $n-1$ words, called the \emph{context}, that is: compute $\pr(w_n | w_1^{n-1})$ for all $w_1^n \in \mathcal{S}$.
Using informal words, we would say that the goal is to \emph{predict} the ``next'' word after a given context.

In what follows, let us indicate with $w_{i}^{j}$ the \emph{sequence} of words $w_i \cdots w_j$, for any $1 \leq i \leq j$, that is equal to $\varepsilon$, the \emph{empty} string, whenever $i < j < 0$.

The conditional probability $\pr(w_n | w_1^{n-1})$ is equal to $\prod_{k=1}^{n}\pr(w_k | w_1^{k-1})$, i.e., all contexts of length $1, 2, \ldots, n-1$ contribute to the final computed value.
Therefore, computing such probability \emph{exactly} is inefficient in both time and memory requirements when $n$ is large.
To make this task feasible and \emph{efficient}, $n$-gram language models are adopted.
%An \emph{$n$-gram} is a sequence of $n$ tokens. A token can be either a single character or a word, the latter intended as a sequence of characters delimited by a special symbol, e.g., a whitespace character.
Unless otherwise specified, throughout the paper we consider datasets of $n$-grams consisting of words.
Since we impose that $1 \leq n \leq N$, where $N$ is a small constant (e.g., $N=5$), dealing with strings of this form permits to work with a context of \emph{at most} $N-1$ preceding words.
This ultimately implies that the aforementioned probability $\pr(w_n | w_1^{n-1}) = \prod_{k=1}^{n}\pr(w_k | w_1^{k-1})$ can be approximated with $\prod_{k=1}^{n}\pr(w_k | w_{k-N-1}^{k-1})$.

Now, the way each $N$-gram probability $\pr(w_k | w_{k-N-1}^{k-1})$ is computed depends on the chosen language model.

\begin{figure}[t]
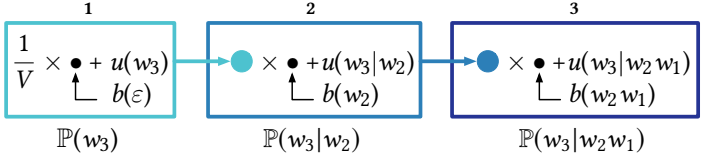

\centering
\includegraphics[scale=0.28]{{{figures/KN-prob}}}
\caption{The Kneser-Ney interpolated probabilities for a 3-gram, calculated in a bottom-up fashion, from left (1-gram) to right (3-gram).
\label{fig:KN-prob}}
\end{figure}

\subsection{Modified Kneser-Ney smoothing}\label{subsec:kneser-ney}
Several models have been proposed in the literature, such as Laplace, Good-Turing, Katz, Jelinek-Mercer, Witten-Bell and Kneser-Ney (see~\cite{CG96,CG99} and references therein for a complete description and comparison).
For an $n$-gram backoff-smoothed language model, the probability of $w_n$ with context $w_1^{n-1}$ is computed according to the following recursive equation
\[
\pr(w_n | w_1^{n-1}) = 
\begin{cases} 
      \pr(w_n | w_1^{n-1}) & \text{if $n$-gram } w_1^n \in \mathcal{S} \\
      b(w_1^{n-1}) \times \pr(w_n | w_2^{n-1}) & \text{otherwise}
 \end{cases}
\]
that is: if the model has enough information we use the full distribution $\pr(w_n | w_1^{n-1})$, otherwise we \emph{backoff} to the lower-order distribution $\pr(w_n | w_2^{n-1})$ with penalty $b(w_1^{n-1})$.

Among these many models, the \emph{modified} version of Kneser-Ney smoothing~\cite{KN95}, introduced by~\citet*{CG96}, was shown to have the best performance in terms of perplexity score.
As already mentioned in Section~\ref{sec:intro}, modified Kneser-Ney is the de-facto choice for language modelling and all major software packages support it.
For these reasons, we adopt Kneser-Ney in this work.

Under this model, the conditional probability $\pr(w_n | w_1^{n-1})$ is computed as
\begin{equation}\label{eq:kn_prob}
\pr(w_n | w_1^{n-1}) = u(w_n | w_1^{n-1}) + b(w_1^{n-1}) \times \pr(w_n | w_2^{n-1})
\end{equation}
Refer to Figure~\ref{fig:KN-prob} for an example of a 3-gram probability computation.
In particular, notice that all lower-order probabilities are \emph{interpolated} together and $u(w_n | w_1^{n-1})$ and $b(w_1^{n-1})$ are, respectively, the normalized probability and context backoff for $n$-gram $w_1^n$
\begin{align}
& u(w_n | w_1^{n-1}) = \frac{a(w_1^n) - D_n(a(w_1^n))}{\sum_x a(w_1^{n-1}x)} \label{eq:norm} \\
& b(w_1^{n-1}) = \frac{\sum_{k=1}^2 D_n(k) \times N_k(w_1^{n-1}\bullet) + D_n(3) \times N_{3+}(w_1^{n-1}\bullet)}{\sum_x a(w_1^{n-1}x)} \label{eq:backoff}
\end{align}

We now explain the quantities used in the above equations.
The quantity $a(w_1^n)$ is called the \emph{modified count} for the $n$-gram $w_1^n$ and it is equal to: either the raw occurrence count $c(w_1^n)$ of $w_1^n$ in the text when $n = N$ or to $|\{x : x w_1^n\}|$ when $n < N$, that is the number of distinct words to the left of $w_1^n$ (also called the \emph{left extensions} of $w_1^n$).
The quantity $N_{k}(w_1^{n-1}\bullet) = |\{x : a(w_1^{n-1}x) = k\}|$ represents the number of $n$-grams having context $w_1^{n-1}$ and modified count equal to $k$, whereas $N_{3+}(w_1^{n-1}\bullet)$ is equal to $|\{x : a(w_1^{n-1}x) \geq 3\}|$, i.e., the number of $n$-grams having modified count greater than or equal to 3.

The recursion shown in Equation~\ref{eq:kn_prob} terminates when unigrams are interpolated with the probability of the \emph{unknown word} which is uniformly distributed by assumption:
$\pr(w_n) = u(w_n) + b(\varepsilon) \times \frac{1}{V}$,
where $V$ denotes the size of the vocabulary, i.e., the number of distinct words appearing in the textual collection used for estimating the language model.
Notice that $b(\varepsilon) / V$ is a constant quantity that depends on the used textual collection.

Lastly, following~\cite{CG96,CG99}, closed-form discounts $D_n(k)$ are computed according to
\begin{equation}\label{eq:discount}
D_n(k) =
\begin{cases}
0 & \text{if $k = 0$} \\
k - (k+1) \times \frac{t_{n,1}t_{n,k+1}}{(t_{n,1} + 2t_{n,2})t_{n,k}} & \text{if $k = 1,2,3$} \\
D_n(3) & \text{otherwise}
\end{cases}
\end{equation}
with the smoothing statistic $t_{n,k}$ representing the total number of $n$-grams in the corpus with modified count $k$, i.e., $t_{n,k} = |\{w_1^n : a(w_1^n) = k\}|$ for $k = 1, 2, 3$ and $4$.

\section{Related Work}\label{sec:related}

In this section we review the solutions proposed in the literature for the two problems that we address in the paper, i.e., indexing and estimation respectively.

\subsection{Indexing}\label{subsec:indexing:related}
We first discuss the classic data structures used to represent large $n$-gram datasets, highlighting the advantages/disadvantages of these approaches in relation to the structural properties that $n$-gram datasets exhibit.
Next, we consider how these approaches have been adopted by different proposals in the literature.
Two different data structures are mostly used to store large and sparse $n$-gram datasets: \emph{tries}~\cite{trie} and \emph{hash tables}~\cite{hashing}.

%\parag{Tries.}
\return
A trie is a tree data structure devised for efficient indexing and search of string dictionaries, in which the common prefixes shared by the strings are represented once to achieve compact storage.
This property makes this data structure useful for storing the $n$-gram strings in compressed space.
In this case, each constituent word of an $n$-gram is associated a node in the trie and different $n$-grams correspond to different root-to-leaf paths. These paths must be traversed to resolve a query, which retrieves the string itself or an associated satellite value, e.g., a frequency count.

Conceptually, a trie implementation has to store a \emph{triplet} for any node: the associated word, satellite value and a pointer to each child node.
As $n$ is typically very small and each node has many children, tries are shallow and wide. Therefore, these are implemented as a collection of (few) sorted arrays: for each level of the trie, a separate array is built to contain all the triplets for that level, sorted by the words. In this implementation, a pair of adjacent pointers indicates the sub-array listing all the children for a word, which can be inspected by binary search.

%\parag{Hash tables.}
\return
Hashing is another way to implement associative arrays: for each value of $n$ from 1 to $N$ a separate hash table stores all the $n$-grams. At the location indicated by the hash function we store a fingerprint value to lower the probability of a false positive
(typically the $4$ or $8$-byte hash of the $n$-gram itself) and the satellite data for the $n$-gram.

This data structure permits to access the specified $n$-gram data in expected constant time.
Open addressing with linear probing is usually preferred over chaining for its better locality of accesses.

\return
Tries are usually designed for space-efficiency as the formed sorted arrays are highly compressible. However, retrieval for the value of an $n$-gram involves exactly $n$ searches in the constituent arrays.
Conversely, hashing is designed for speed but sacrifices space-efficiency since its keys, along with their (usually expensive) fingerprint values, are randomly distributed and, therefore, incompressible.
Furthermore, hashing is a randomized solution, i.e., there is a non-null probability of retrieving a frequency count for an $n$-gram \emph{not} really belonging to the indexed corpus (false positive). Such probability equals $2^{-b}$, where $b$ indicates the number of bits dedicated to the fingerprint values: larger values of $b$ yield a smaller probability of false positive but also increase the space of the data structure.

\return
The paper by Pauls and Klein~\cite{PK11} proposes trie-based data structures with nodes represented via sorted arrays or via hash tables with linear probing.
The sorted arrays are compressed using a variable-length block encoding, consisting in the following steps: (1) a configurable radix $r = 2^k$ is chosen; (2) the number of digits, $d$, needed to represents a number in base $r$ is written in unary; (3) the $d$ digits are written in $dk$ bits.
To preserve the property of looking up a record by binary search, each sorted array is divided into blocks of $128$ bytes.
The encoding is used to compress words, pointers and the positions that frequency counts take in a unique-value array that collect all distinct counts.
%This unique-value array is very small and its storage negligible.
The hash-based variant is faster than the sorted array variant, but requires extra table allocation space to avoid excessive collisions.
%Instead of the frequency count itself, the position of the count in a unique-value array that collects all distinct counts is stored. 
%It is convenient to store the rank of a count instead of the count itself because the distribution of frequency counts is highly repetitive: the number of distinct counts is far less than the number of $n$-grams, allowing to represent a count with a number of bits equal to the binary logarithm of the number of distinct counts rather than the maximum count value.
%Moreover, frequency counts follow a Zipfian distribution too, i.e., far more $n$-grams have a very low count than a high one.
%For this reason, small values are expected to appear more frequently than large ones.

Heafield~\cite{Heafield11} improves the sorted array trie implementation with some optimizations.
The keys in the arrays are replaced by their hashes and sorted, so that these are uniformly distributed over their ranges. Now finding a word ID in a trie level of size $m$ can be done in $\Oh(\log\log m)$\footnote{Unless otherwise specified, all logarithms are in base 2 and $\log x = \log_2 x$ for any $x > 0$.} with high probability by using \emph{interpolation} search~\cite{DTP04}.
Records in each sorted array are minimally sized at the bit level, improving the memory consumption over~\cite{PK11}.
Pointers are compressed using the integer compressor devised in~\cite{RW03}.
Values can also be quantized using the \emph{binning} method~\cite{BF06} that sorts the values, divides them into equally-sized bins and then elects the average value of the bin as the representative of the bin. The number of chosen quantization bits directly controls the number of created bins and, hence, the trade-off between space and accuracy.

Talbot and Osborne~\cite{TO07} use Bloom filters~\cite{Bloom70} with lossy quantization of frequency counts to achieve small memory footprint. In particular, the raw frequency count $f_g$ of the $n$-gram $g$ is quantized using a logarithmic codebook, i.e., $\widetilde{f}_g = 1+ \log_b f_g$. The scale is determined by the base $b$ of the logarithm: in the implementation $b$ is set to $2^{1/v}$, where $v$ is the quantization range used by the model, e.g., $v = 8$.
Given the quantized count $\widetilde{f}_g$ of $g$, a Bloom filter is trained by entering composite events into the filter, represented by $g$ with an appended integer value $j$, which is incremented from 1 to $\widetilde{f}_g$.
In order to retrieve $\widetilde{f}_g$ at query time, the filter is queried with a $1$ appended to $g$. This event is hashed using the $k$ hash functions of the filter: if all of them test positive, then the count is incremented and the process repeated.
The procedure terminates as soon as any of the $k$ hash functions hits a $\textsf{0}$ and the previous count is reported (after conversion to a linear count)
This procedure avoids a space requirement for the counts proportional to the number of grams in the corpus because only the codebook needs to be stored.
The one-sided error of the filter and the training scheme ensure that the actual quantized count cannot be larger than the reported value. As the counts are quantized using a logarithmic-scaled codebook, the count will be incremented only a small number of times.

The use of the succinct encoding \textsf{LOUDS} (Level-Order Unary-Degree Sequence)~\cite{Jacobson89} is advocated in the work by~\citet{Watanabe09} to implicitly represent the trie nodes. In particular, the pointers for a trie of $m$ nodes are encoded using a bitvector of $2m + 1$ bits.
Bit-level searches on such bitvector allow forward/backward navigation of the trie structure.
Words and frequency counts are compressed using Variable-Byte encoding~\cite{thiel1972program,Salomon07}, with an additional bitvector used to indicate the boundaries of such byte sequences as to support random access to each element.
%The paper also discusses the use of block-wise compression (basically \textsf{gzip} on blocks of $8$ KB) though it is not used in the implementation for time efficiency reasons.
~\citet{shareghi2015compact,shareghi2016fast} also consider the usage of succinct data structures to represent \emph{suffix trees} that can be used to compute Kneser-Ney probabilities on-the-fly. Experimental results indicate that the method is practical for large-scale language modeling although significantly slower to query than leading toolkits for language modeling~\cite{Heafield11}.

The problem of representing trie-based storage for general-purpose string dictionaries is among one of the most studied in computer science, with many and different solutions available~\cite{HZW02,NBYST02,CM96}.
It goes without saying that, given the properties that $n$-gram datasets exhibit, generic trie implementations are \emph{not} suitable for their efficient treatment.
However, comparing with the performance of such implementations gives useful insights about the performance gap with respect to a general solution.
We mention $\mari$~\cite{Yata11} as the best and practical general-purpose trie implementation. The main idea is to use \textsf{Patricia} tries~\cite{Morrison68} to recursively represent the nodes of a \textsf{Patricia} trie. This clearly comes with a space/time trade-off: the more levels of recursion are used, the greater the space saving but also the higher the retrieval time.

\subsection{Estimation}\label{subsec:estimation:pre}
In this subsection we first discuss the related work concerning the estimation of language models, then we describe the state-of-the-art algorithm devised by~\citet{HPCK13}.

\return
The use of the \textsf{Map+Reduce} paradigm for the problem has been advocated in~\cite{Brantz07}. As reported in the paper, estimation involved hundreds of machines for a few days. Our work does not consider distributed computations, rather it shows how to let the estimation process scale well on the cores of a single target machine.
~\citet{MSRLM} (\textsf{MSRLM}) also considered estimation on a single machine, using a parallel merge sort implementation. However, part of the estimation process is delayed until query-time: while this allows to save some resources during estimation, it also imposes a significant burden during the most efficiency-demanding use of language models which is query processing~\cite{CG99,Heafield11}.
We, instead, prefer to follow the approach of~\cite{HPCK13} that performs all the steps of estimation as to permit the building of an efficient, static, compressed index over the computed model.

The works done by \citet{SRILM} (\textsf{SRILM}), \citet{IRSTLM} (\textsf{IRSTLM}), \citet{PK11} ($\berk$) and \citet{Watanabe09} ($\expg$) build Kneser-Ney language models in internal memory without resorting on sophisticated software optimizations and data compression techniques: as a result, are not able to scale to the dimensions we consider in this work.

~\citet{HPCK13} ($\ken$) contributed an estimation algorithm involving three steps of sorting in external memory.
Their solution, referred to as the $\ts$ algorithm in the following, significantly outperforms the approaches that we have mentioned above, making it the state-of-art solution to the problem.
%Therefore, we describe this algorithm in the following.
%Indeed, as the authors reported in the paper~\cite{HPCK13}, their algorithm takes, on average, as low as $20\%$ of the CPU and $10\%$ of the RAM of both \textsf{SRILM} and \textsf{IRSTLM}.
Specifically, it takes takes $25.4\%$ and $7.7\%$ of, respectively, CPU time and RAM of \textsf{SRILM}; $16.4\%$ and $16.6\%$ of CPU and RAM of \textsf{IRSTLM}~\cite{HPCK13}.

The recent approach by~\citet{shareghi2015compact} resorts on compressed suffix trees to compute the Kneser-Ney probabilities on the fly.
The experimental analysis reported in the paper compares against \textsf{SRILM} and shows that such approach is comparable in building time with \textsf{SRILM} indexes but several orders of magnitude (e.g., $1000\times$) slower to query.
In~\cite{shareghi2016fast} the same authors improved over their previous work~\cite{shareghi2015compact} by pre-computing some modified counts to speed up the on-the-fly calculation of the Kneser-Ney probabilities.
Although pre-computing allows for significant improvement at query time (by up to $2500\times$ faster than the previous solution) at the price of a larger index construction time ($70\%$ more time), the resulting language model is still $5\times$ slower than $\ken$.

\return
For the reasons discussed above, we aim at improving upon the I/O efficiency of the $\ts$ approach by~\citet{HPCK13} ($\ken$) that we describe in details in the following.

\subsubsection{The 3-Sort algorithm}\label{3-sort}
%Now we review the algorithm devised by~\citet{HPCK13} since our work aims at improving its I/O efficiency.
%
%\return
During the estimation process, we deal with the following assumptions:
\begin{enumerate}[leftmargin=*]
\item the uncompressed $n$-gram strings with associated satellite values, $1 \leq n \leq N$ do \emph{not} fit in internal memory and we necessarily need to rely on disk usage;
\item the estimate is performed \emph{without} pruning, thus the minimum occurrence count for an $n$-gram is 1;
\item the compressed index built over the $n$-gram strings
%(i.e., the trie presented in Section~\ref{subsec:indexing:ef-trie}) 
\emph{must} reside in internal memory to allow fast query processing (e.g., for perplexity-score computations and machine translation).
\end{enumerate}

\begin{figure}[t]
%\begin{wrapfigure}{l}{0.4\textwidth}
    \centering
    \subfloat[]{
    \includegraphics[scale=0.53]{{{figures/suffix-order}}}
    \label{fig:suffix-order}
    }
    \hspace{0.5cm}
    \subfloat[]{
    \includegraphics[scale=0.53]{{{figures/context-order}}}
    \label{fig:context-order}
    }

    \caption{A block of 12 5-grams sorted in \emph{suffix} order (a) and sorted in \emph{context} order (b).}
    \label{fig:order}
%\end{wrapfigure}
\end{figure}

Since the sorted orders defined over a set of $N$-grams are central to the description of the algorithm we are going to consider, we define them as follows.
Consider a set of $N$-grams\footnote{Throughout the paper, whenever we need to show some examples, we consider an $n$-gram as consisting of $n$ capital letters rather than words.} as the one illustrated in Figure~\ref{fig:order}.
The set is put into sorted order by sorting the $N$-grams on their words, as considered in a specific order.
If the specific order is $N,N-1,\ldots,1$, i.e., we sort the $N$-grams from their last word up to the first, then the block is \emph{suffix}-sorted: the last word is primary (Figure~\ref{fig:suffix-order}).
If the considered order is $N-1,N-2,\ldots,1,N$, then the block is \emph{context}-sorted: the penultimate word is primary (Figure~\ref{fig:context-order}).

\return
As an overview, the algorithm consists in four streaming passes over the data that we are going to detail next: (1) counting; (2) adjusting counts; (3) normalization; (4) interpolation and joining.
Since all $n$-grams, $1 < n \leq N$, are sorted between these steps in the next-step desired order, thus \emph{three} times in total, we refer to this approach as the $\ts$ algorithm.

\begin{enumerate}[leftmargin=*]
%\subsubsection{Counting}\label{subsec:estimation:3_sort:counting}
\return
\item
\textbf{Counting.}
The first step computes the unpruned occurrence counts $c(w_1^N)$ for all the distinct $N$-grams in the text (with order exactly $N$) by streaming through the textual corpus, using a a window of size $N$ words that slides by one word at a time.
Lower-order $n$-grams are not counted since raw occurrence counts for $N$-grams are sufficient to derive smoothing statistics.
In particular, $N$-gram tokens are replaced with 4-byte vocabulary identifiers and unigram strings are written to disk as plain text. Their 8-byte Murmur
%\footnote{\url{https://github.com/aappleby/smhasher}}
hash is retained in internal memory.
The occurrence counts, represented as 8-byte numbers, are accumulated in an open-addressing hash table with linear probing: the counts are finally written to disk in a \emph{suffix-sorted} block as records of the form $\langle w_1^N, c(w_1^N) \rangle$ whenever the table reaches a specified amount of internal memory.

%\subsubsection{Adjusting}\label{subsec:estimation:3-sort:adjusting}
\return
\item
\textbf{Adjusting counts.}
All blocks sorted in suffix order are merged together in a single block $B_N$.
This step aims at computing the modified counts $a(w_1^n)$ for the $n$-grams $w_1^n$ that is equal to $|\{x : x w_1^n\}|$, which is the number of distinct words to the left of $w_1^n$.

By streaming through $B_N$ sorted in \emph{suffix} order it is sufficient to compare consecutive entries to decide whether to write the record $\langle w_1^n, a(w_1^n) \rangle$ to a new block $B_n$ or increment the currently computed $a(w_1^n)$. During the same pass, smoothing statistics $t_{n,k}$ are collected and discount coefficients $D_n(k)$ are calculated as in Equation \ref{eq:discount}.

%\subsubsection{Normalization}\label{subsec:estimation:3-sort:normalization}
\return
\item
\textbf{Normalization.}
This step computes normalized probabilities and backoffs according to, respectively, Equation \ref{eq:norm} and \ref{eq:backoff}.
For such purpose, the blocks $B_n$, $1 < n \leq N$, produced during the previous step, are sorted in context order such that, for each context $w_1^{n-1}$, the entries $w_1^{n-1}x$ are consecutive.
Also in this case, a streaming pass through each block $B_n$ suffices to emit records of the form $\langle w_1^n, u(w_n | w_1^{n-1}), b(w_1^{n-1}) \rangle$. The information stored in the record (refer to Figure~\ref{fig:KN-prob}) is one needed to perform interpolation.
The computed backoffs are saved twice on disk, also as bare values without keys, one file per order $1 \leq n < N$ to facilitate the next step of interpolation and joining.

%\subsubsection{Interpolation and joining}\label{subsec:estimation:3-sort:interpolation-joining}
\return
\item
\textbf{Interpolation and joining.}
The last streaming step performs interpolation of all orders to compute the final Kneser-Ney probability as in equation \ref{eq:kn_prob}.
The blocks $B_n$ are sorted again in suffix order so that $\pr(w_n)$ is computed before it is needed to compute $\pr(w_n | w_{n-1})$, which in turn is computed before $\pr(w_n | w_{n-2}w_{n-1})$, and so on.
Figure~\ref{fig:KN-prob} offers a pictorial representation of this bottom-up process for a 3-gram.
Note that the backoffs for the contexts that are needed for interpolation were saved in-line with the string $w_1^n$ during the previous step.
Also, note that since normalization streamed through the blocks sorted in context order, the backoffs were saved to disk in suffix order.
Therefore, during this step the two quantities $\pr(w_n | w_1^{n-1})$ and $b(w_1^n)$ are joined together, for $1 \leq n < N$ ($N$-grams do not have backoff).
\end{enumerate}

\section{Compressed Indexes}\label{sec:indexing}
The problem we tackle in this section of the paper is the one of representing in compressed space a dataset of $n$-gram strings and their associated values, being either frequency counts (integers) or probabilities (floating points).
Given an $n$-gram string, the compressed data structure should allow fast random access to the corresponding associated value by means of the operation $\lk$.

\subsection{Elias-Fano tries}\label{subsec:indexing:ef-trie}
In this subsection we present a compressed trie data structure based on the \emph{Elias-Fano} representation~\cite{Elias74,Fano71} of monotone integer sequences for its efficient random access and search operations.
As we will see, the constant-time random access of Elias-Fano makes it the right choice for the encoding of the sorted-array trie levels, given that we (fundamentally) need to randomly access the sub-array pointed to by a pair of pointers.
Such pair is retrieved in constant time too.
Now every access performed by binary search takes $\Oh(1)$ \emph{without} requiring any block decompression, differently from other strategies~\cite{PK11}.

We also introduce a novel technique to lower the memory footprint of the trie levels by losslessly reducing the entity of their constituent integers. This reduction is achieved by mapping a word identifier (ID in the following) \emph{conditionally} to its context of fixed length $k$, i.e., its $k$ preceding words.

\subsubsection{Data structure}\label{subsec:indexing:core}
%This subsection contains the core description of the compressed trie data structure: we dedicate one paragraph to each of its main building components, i.e., how the grams, satellite data and pointers are represented; how searches are implemented.

\begin{figure}
    \centering
    \subfloat[]{
    \includegraphics[scale=0.43]{{{figures/trigrams-trie}}}
    \label{fig:trigrams-trie}
    }
    \subfloat[]{
    \includegraphics[scale=0.43]{{{figures/trigrams-arrays-not-monotone}}}
    \label{fig:trigrams-arrays-not-monotone}
    }
    \subfloat[]{
    \includegraphics[scale=0.43]{{{figures/trigrams-arrays-monotone}}}
    \label{fig:trigrams-arrays-monotone}
    }
    \caption{In (\textsf{a}) we show an example of a trie of order $3$, representing the set of grams $\{$\textsf{A}, \textsf{AA}, \textsf{AAC}, \textsf{AC}, \textsf{B}, \textsf{BB}, \textsf{BBC}, \textsf{BBD}, \textsf{BC}, \textsf{BCD}, \textsf{BD}, \textsf{C}, \textsf{CA}, \textsf{CD}, \textsf{D}, \textsf{DB}, \textsf{DBB}, \textsf{DBC}, \textsf{DDD}$\}$.
    In (\textsf{b}) we see the sorted-array representation of the trie, where each vocabulary token is assigned a distinct integer ID. Lastly, in (\textsf{c}), we show the final representation of the trie where each sorted array has been transformed in a monotone sequence by computing the prefix sums of the ranges marked with the thick bars in (\textsf{b}).
    The shaded arrays represent the pointers.}
    \label{fig:trigrams}
    \vspace{-0.5cm}
\end{figure}

As it is standard, a unique integer ID is assigned to each distinct word to form the vocabulary of the indexed $n$-gram corpus.
Such vocabulary is implemented using a hash data structure that stores for each unigram its ID in order to retrieve it when needed in $\Oh(1)$.
If we sort the $n$-grams following the token-ID order, we have that all the successors of gram $w_1^{n-1} = w_1 \cdots w_{n-1}$, i.e., all grams whose prefix is $w_1^{n-1}$, form a strictly increasing integer sequence.
For example, suppose we have the unigrams $\{$\textsf{A}, \textsf{B}, \textsf{C}, \textsf{D}$\}$, which are assigned IDs $\{ 0, 1, 2, 3 \}$ respectively.
Now consider the bigrams $\{$\textsf{AA}, \textsf{AC}, \textsf{BB}, \textsf{BC}, \textsf{BD}, \textsf{CA}, \textsf{CD}, \textsf{DB}, \textsf{DD}$\}$ sorted by IDs. The sequence of the successors of \textsf{A}, referred to as the \emph{range} of \textsf{A} in this paper, is $\langle \textsf{A}, \textsf{C} \rangle$, i.e., $\langle 0, 2 \rangle$, because \textsf{A} and \textsf{C} are prefixes by \textsf{A} to form the bigrams \textsf{AA} and \textsf{AC}; the sequence of the successors of \textsf{B}, is $\langle$\textsf{B}, \textsf{C}, \textsf{D}$\rangle$, i.e., $\langle 1, 2, 3 \rangle$ and so on.
Concatenating all such ranges, we obtain the integer sequence $\langle 0, 2 | 1, 2, 3 | 0, 3 | 1, 3 \rangle$ that we can see in Figure~\ref{fig:trigrams-arrays-not-monotone} (the thick vertical bars, depicted in dark blue in Figure~\ref{fig:trigrams-arrays-not-monotone}, are not really part of the sequence: they are shown to better highlight the different ranges).
In order to distinguish the successors of an $n$-gram from others, we also maintain where each range begins in a monotone integer sequence of pointers.
In our example, the sequence of pointers is $\langle 0, 2, 5, 7, 9 \rangle$ (we also store a final dummy pointer to be able to obtain the last range length by taking the difference between the last and previous pointer).
The ID assigned to a unigram is also used as the position at which we read the unigram pointer in the unigrams pointer sequence.

Therefore, apart from unigrams that are stored in a hash table, each level of the trie is composed by \emph{two} integer sequences: one for the representation of the gram-IDs, the other for the pointers.
Figure~\ref{fig:trigrams} shows a graphical representation of what we described.

\return
We have therefore reduced the problem of representing a trie to the problem of compressing (a few) integer sequences.
While many integer compressors are available in the literature,
in this work we adopt Elias-Fano ({\EF})~\cite{Elias74,Fano71}, along with its \emph{partitioned} variant ({\PEF})~\cite{OV14}, which has been applied to inverted index compression showing an excellent space/time trade-off~\cite{Vigna13,OV14,PV17}. %thanks to its powerful search capabilities.

We now quickly state the salient features of this elegant integer encoding
and we point the reader to the recent survey by~\citet*{EBDT2018} for a more detailed description.

Elias-Fano encodes a monotonically increasing sequence $S(m,u)$ of $m$ positive integers drawn from a universe of size $u$ in
$\EF(\seq(m,u)) \leq m \lceil \log \frac{u}{m} \rceil + 2m$ bits
and it permits to randomly access an integer in constant time, \emph{without} decompressing the whole sequence. Again, see the survey in~\cite{EBDT2018} for a description of the algorithm for random access and references therein.
The \emph{partitioned} Elias-Fano variant, proposed by~\citet*{OV14}, splits the sequence into variable-sized partitions to enhance compression effectiveness.
%Each sequence is represented by two levels: the first level stores the endpoints of the partitions; the second level is the compressed representation of the partitions.
Clearly, the partitioned sequence organization introduces a level of indirection when resolving a random access, because a first search must be spent in the first level to identify the block in which the searched integer is located. We will return to and stress this point in the experimental Section~\ref{subsec:indexing:exp}.

\parag{Gram-ID sequences and pointers}
While the sequences of pointers are monotonically increasing by construction and, therefore, immediately Elias-Fano encodable, the gram-ID sequences may not be, as we can see from Figure~\ref{fig:trigrams-arrays-not-monotone}. However, a gram-ID sequence can be transformed into a \emph{monotone} one (though not strictly increasing) by taking \emph{range-wise} prefix sums:
to the values of a range we add the last prefix sum (initially equal to $0$).
Then, our example sequence $\langle 0, 2 | 1, 2, 3 | 0, 3 | 1, 3 \rangle$ becomes $\langle 0, 2 | 3, 4, 5 | 5, 8 | 9, 11 \rangle$.
The last prefix sum is initially $0$, therefore the range of \textsf{A} remains the same, i.e., $\langle 0, 2 \rangle$. Now the last prefix sum is $2$, so we sum $2$ to the values in the range of \textsf{B}, yielding $\langle 3, 4, 5 \rangle$.
Now the last prefix sum is $5$, so we sum $5$ to the values in the range of \textsf{C}, yielding $\langle 5, 8 \rangle$.
Finally, the last prefix sum is $8$, therefore we sum $8$ to the values in the range of \textsf{D}, obtaining $\langle 9, 11 \rangle$.
The final trie resulting from this transformation is shown in Figure~\ref{fig:trigrams-arrays-monotone}.

In particular, if we sort the vocabulary IDs in decreasing order of occurrence, we make small IDs appear more often than large ones and this is highly beneficial for the growth of the universe $u$ and, hence, for Elias-Fano whose space occupancy critically depends on it.
We emphasize this point again: for each unigram in the vocabulary we count the number of times it appears in all gram-ID sequences and assign IDs to vocabulary tokens in decreasing order of occurrence\footnote{Note that the number of occurrences of an $n$-gram can be different than its frequency count as reported in the dataset. The reason is that such datasets often do not include the $n$-grams appearing less than a predefined frequency threshold.}.

\parag{Frequency counts}
To represent the frequency counts, we use the \emph{unique-value array} technique, i.e., each count is represented by its \emph{index} in an array $C[n]$, $1 \leq n \leq N$, that collects all the \emph{distinct} frequency counts for the $n$-grams.
This technique is widely used in data compression whenever the distribution of the represented values is extremely skewed, as it is in our case for the frequency counts of the $n$-grams: relatively few $n$-grams are very frequent while most of them appear only a few times.
As we can better see in Table~\ref{tab:indexing:datasets_stats} (Section~\ref{subsec:indexing:exp}), the number of distinct counts is very small compared to the number of $n$-grams themselves, so the space for the arrays $C[n]$, $1 \leq n \leq N$, is negligible.

Now, each level of the trie, besides the sequences of gram-IDs and pointers, has also to store the sequence made by all the frequency-count indexes.
Unfortunately, this sequence of indexes is not monotone, yet it follows the aforementioned highly repetitive distribution.
To exploit such repetitiveness, we assigned to each index a codeword of variable length.
As similarly done for the gram-IDs, by assigning smaller codewords to more repetitive indexes, we have most indexes encoded in just few bits.
More specifically, starting from $k=1$, we first assign all the $2^k$ codewords of length $k$ before increasing $k$ by $1$ and repeating the process until all indexes have been considered. Therefore, we first assign codewords $0$ and $1$, then codewords $00$, $01$, $10$, $11$, $000$ and so on. All codewords are then concatenated one after the other in a bitvector $B$.

Following~\cite{FN07}, to the $i$-th index we give codeword $c = i + 2 - 2^{\ell_c}$, where $\ell_c = \lfloor \log(i + 2) \rfloor$ is the number of bits dedicated to the codeword $c$. From codeword $c$ and its length $\ell_c$ in bits, we can retrieve $i$ by taking the inverse of the previous formula, i.e., $i = c - 2 + 2^{\ell_c}$. Besides the bitvector for the codewords themselves, we also need to know where each codeword begins and ends.
We can use another bitvector for this purpose, say $L$, that stores a $1$ for the starting position of every codeword. A small additional data structure built on $L$ allows efficient computation of the $\sel_1$ primitive that we use to retrieve $\ell_c$. In fact, $b = \sel_1(i)$ gives us the starting position of the $i$-th codeword. Its length is easily computed by scanning $L$ upward from position $b$ until we hit the next $1$, say in position $e$. Finally, $\ell_c = e - b$ and $c = B[b, e - 1]$.

\return
In conclusion, a trie is conceptually represented by an array of levels, \emph{levels}$[1,N]$, where each \emph{levels}$[n]$ stores, for $1 \leq n \leq N$: the gram-ID sequence \emph{levels}$[n].ids$, the sequence of frequency-count indexes \emph{levels}$[n].indexes$ and the pointer sequence \emph{levels}$[n].pointers$, with the only exceptions of 1-grams and $N$-grams, for which gram-ID and pointer sequences are missing respectively.

%\begin{figure}[htb]
%
%   \subfloat[]{
%       \scalebox{0.9}{
%           \begin{minipage}{.41\textwidth}
%           \input{lookup-no-remapping}
%           \end{minipage}
%       }
%   }
%   \subfloat[]{
%       \scalebox{0.9}{
%           \begin{minipage}{.55\textwidth}
%           \input{search}
%           \end{minipage}
%        }
%   }
%
%  \caption{The \textbf{$\lk$} and \textbf{\textsf{search}} functions. The \textbf{\textsf{find}}($G_k, x, b, e$) function, used in the \textbf{\textsf{search}} pseudo code, finds the integer $x$ in the range $G_k[b,e)$ and returns its position in $G_k$.
%  \label{alg:lookup-no-remapping}}
%\end{figure}

\begin{figure}[t]

    \subfloat[]{
        \scalebox{0.9}{
            \begin{minipage}{.41\textwidth}
            \input{lookup-no-remapping}
            \end{minipage}
        }
    }
    \subfloat[]{
        \scalebox{0.9}{
            \begin{minipage}{.58\textwidth}
            \input{search}
            \end{minipage}
         }
    }

  \caption{The \textbf{$\lk$} and \textbf{\textsf{search}} functions. The \textbf{\textsf{find}}($A, b, e, x$) function, used in the \textbf{\textsf{search}} pseudo code, finds the integer $x$ in the range $A[b,e)$ and returns its position in $A$.
  \label{alg:lookup-no-remapping-search}}
\end{figure}

\parag{Lookup}
We now describe how the $\lk$ operation is implemented, i.e., how to retrieve the frequency count given an $n$-gram $w_1^n$.
The corresponding pseudo code is illustrated in Figure~\ref{alg:lookup-no-remapping-search}.
We first perform $n$ vocabulary lookups to map the $n$-gram tokens into its constituent IDs. We write these IDs into an array $ids[1,n]$ (Lines 2-4 in Figure~\ref{alg:lookup-no-remapping}).
This preliminary query-mapping step takes $\Theta(n)$ because each vocabulary lookup is performed in $O(1)$. 
Now, the search procedure has to locate $ids[i]$ in the $i$-th level of the trie (Lines 3-6 in Figure~\ref{alg:search}), as follows.
If $n = 1$, then our search terminates: at the position $p = ids[1]$ we read the index $i$ = \emph{levels}$[1].indexes[p]$ to finally return $C[1][i]$. If, instead, $n$ is greater than $1$, the position $p$ is used to retrieve the pair of pointers ($b, e$) = (\emph{levels}$[1].pointers[p]$, \emph{levels}$[1].pointers[p + 1]$) in constant time, which delimits the range of IDs in which we have to search for $ids[2]$ in the second level of the trie.
This range is inspected by binary search with the operation \texttt{\textbf{find}}, taking $\Oh(\log (e-b))$ because each access to an Elias-Fano-encoded sequence is performed in constant time.
Now $p$ is updated to be the position in \emph{levels}$[2].ids$ at which $ids[2]$ is found in the range. Again, if $n = 2$, the search terminates by accessing $C[2][i]$ where $i$ is now the index \emph{levels}$[2].indexes[p]$.
If $n$ is greater than $2$, we fetch the pair (\emph{levels}$[2].pointers[p]$, \emph{levels}$[2].pointers[p+1]$) to continue the search of $ids[3]$ in the third level of the trie, and so on.
This search step is repeated for $n-1$ times in total, to finally return the count $C[n][i]$ of $w_1^n$.

\subsubsection{Context-based identifier remapping}\label{subsec:indexing:remapping}
In this subsection we describe a novel technique that lowers the space occupancy of the gram-ID sequences that constitute, as we have seen, the main component of the trie data structure.

The key idea is to map a word $w$ occurring after the context $w_1^k$ to an integer whose value is bounded by the number of words that \emph{follow} such context, and \emph{not} bounded by the total vocabulary size $V$.
Specifically, $w$ is mapped to the position it occupies within its siblings, i.e., the words following the $k$-gram $w_1^k$. We call this technique \emph{context-based identifier remapping} because each ID is re-mapped to the position it takes relatively to a context.

\begin{figure}
    \centering    
    \subfloat[]{
    \includegraphics[scale=0.7]{{{figures/remapping}}}
        \label{fig:remapping}
    }   
    \hspace{0.4cm}
    \subfloat[]{
    \scalebox{0.9}{
    \begin{tabular}[b]{
        c@{\hspace{10pt}}       
        c@{\hspace{10pt}}
        r@{\hspace{2pt}}
        r@{\hspace{8pt}}
        r@{\hspace{2pt}}
        r@{\hspace{8pt}}
        r@{\hspace{2pt}}
        r@{\hspace{8pt}}
    }
    \toprule
    \input{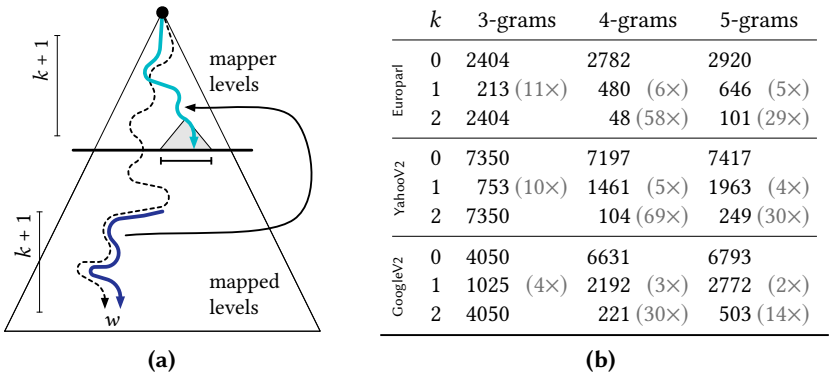}
    \bottomrule
    \end{tabular}
    \label{tab:stats}
    }}
    
    \caption{In (\textsf{a}), we depict the action performed by the context-based identifier remapping strategy. The last word ID $w$ of any sub-path of length $k+1$, e.g., the dark blue one, is replaced with the position it takes within its sibling IDs. These sibling IDs are found at the end (the gray triangle) of the search of $w$ along the \emph{same} path, e.g., the light green one, in the first $k+1$ levels of the trie.
    In (\textsf{b}), we show the effect of the context-based remapping on the average gap (ratio between universe and size) of the gram-ID sequences of the datasets used in the experiments, with context length $k = 0, 1, 2$.}
    \vspace{-0.3cm}
\end{figure}

Figure~\ref{fig:remapping} shows a representation of the action performed by the remapping strategy: the last word ID $w$ of any sub-path of length $k+1$ (e.g., the dark blue one in the figure) is searched along the \emph{same} path occurring in the first $k+1$ levels of the trie (e.g., the light green one in the figure). This can be graphically interpreted as if the dark blue path were projected onto the light green path in order to search $w$ along its sibling IDs that are the ones occurring after the $k$-gram $w_1^k$ (the small dark gray triangle in the figure).
We remark the fact that this projection is \emph{always possible}, i.e., we are guaranteed to find any sub-path of length $k+1$ in the first $k+1$ levels of the trie, because of the sliding-window extraction process described in Section~\ref{sec:background}.
Figure~\ref{fig:remapping} also highlights that using a context of length $k$ will partition the levels of the trie into two categories: the so-called \emph{mapper} levels and the \emph{mapped} levels. The first $k+1$ levels of trie act, in fact, as a mapper structure whose role is to map any word ID through searches; all the other $N - k - 1$ levels are the ones formed by the remapped IDs.

The salient feature of the strategy is that it takes full advantage of the $n$-gram model represented by the trie structure itself in that it does \emph{not} need any redundancy (e.g., an additional data structure) to perform the mapping of IDs, because these are mapped by means of searches in the first $k+1$ levels of the trie.
The strategy also allows a great deal of flexibility in that we can choose the length $k$ of the context. 
In general, for an $n$-gram dataset comprising all $n$-grams for $n = 1, \ldots, N$ with $N \geq 2$, we can choose between $N-2$ distinct context lengths $k$, i.e., $1 \leq k \leq N - 2$.
Clearly, the greater the context length we use, the smaller the remapped IDs will be but the searches will take longer. The choice of the proper context length to use should take into account the characteristics of the $n$-gram dataset; in particular the \emph{number of $n$-grams} per order.

In what follows we explain \emph{why} the introduced remapping strategy offers a valuable contribution to the overall space reduction of the trie data structure, throughout some didactic and real examples.
As we will see in the experimental Section~\ref{subsec:indexing:exp}, the dataset vocabulary can contain several million tokens, whereas the number of words that naturally occur after another is typically very small.
Even in the case of stopwords, such as ``the'' or ``are'', the number of words that can follow is \emph{far less} than the whole number of distinct words for \emph{any} (reasonably large) $n$-gram dataset.
This ultimately means that the remapped integers forming the gram-ID sequences of the trie will be \emph{much smaller} than the original ones, which can indeed range from $0$ to $V-1$.
Lowering the values of the integers clearly helps in reducing the memory footprint of the levels of the trie because \emph{any} integer compressor takes advantage of encoding smaller integers, since fewer bits are needed for their representation~\cite{MS00,OV14}.
In our case the gram-ID sequences are encoded with Elias-Fano: from Section~\ref{subsec:indexing:core}, we know that Elias-Fano spends $\lceil\log \frac{u}{m}\rceil + 2$ bits per integer, thus a number of bits proportional to the average gap $u/m$ between its values.
The remapping strategy reduces the universe $u$ of representation, thus lowering the average gap and space of the sequence.

\begin{figure}
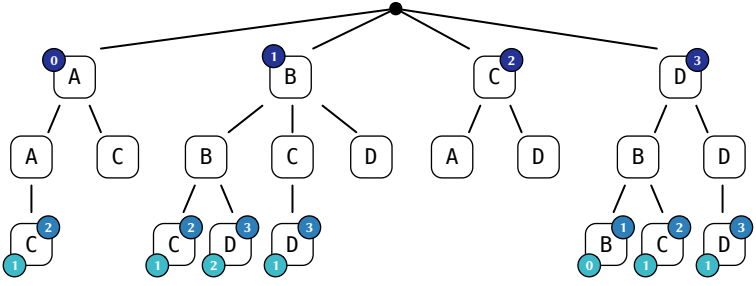

    \centering
    \includegraphics[scale=0.50]{{{figures/trie2}}}
    \caption{Example of a trie of order $3$, representing the set of grams $\{$\textsf{A}, \textsf{AA}, \textsf{AAC}, \textsf{AC}, \textsf{B}, \textsf{BB}, \textsf{BBC}, \textsf{BBD}, \textsf{BC}, \textsf{BCD}, \textsf{BD}, \textsf{CA}, \textsf{CD}, \textsf{DB}, \textsf{DBB}, \textsf{DBC}, \textsf{DDD}$\}$. From up to down, we show: the vocabulary IDs in darkest blue, level-$3$ IDs in blue and finally the light green IDs (the ones in the left-down corner) are derived by applying a context-based remapping with context length $1$.}
    \label{fig:trie}    
\end{figure}

This effect of the strategy is illustrated in Figure~\ref{tab:stats} that shows how the average gap of the gram-ID sequences of the datasets we used in the experiments (see also Table~\ref{tab:indexing:datasets_stats}, Section~\ref{subsec:indexing:exp}) is affected by the context-based remapping.
As unigrams and bigrams constitute the mapper levels, these are kept unmapped: we show the statistics for the mapped levels, i.e., the third, fourth and fifth, of a trie of order $5$ built from the $n$-grams of the datasets.
For each dataset we did the experiment for context lengths $0$, $1$ and $2$.
As we can see by considering $\euro$, the technique with a context of length $1$ achieves an average reduction of $7.2\times$ (up to $11.3\times$ on tri-grams).
With a context of length $2$, instead, we obtain an average reduction of $43.4\times$ (up to $58\times$ on $4$-grams).
Very similar considerations and numbers hold for the $\yahoo$ dataset as well.
The reduction on the $\google$ dataset is less dramatic instead, being on average of $3\times$ with a context of length $1$ and of $16.75\times$ with a context of length $2$.

\parag{Lookup}
The described remapping strategy comes with an overhead at query time because the lookup algorithm illustrated in Figure~\ref{alg:lookup-no-remapping-search} must map a default vocabulary ID to its remapped ID, before it can be searched in the proper trie level.

\begin{wrapfigure}{l}{.45\textwidth}
  \begin{minipage}{.5\textwidth}
    \scalebox{0.9}{\input{lookup-remapping}}
  \end{minipage}
  \caption{The \textbf{$\lk$} function with context-based remapping of order $k$.
  \label{alg:lookup-remapping}}
\end{wrapfigure}

Specifically, if the remapping strategy is applied with a context of length $k$, it involves $k \times (N-k-1)$ additional searches in the trie levels. As an example, by looking at Figure \ref{fig:trie}, before searching the mapped ID $1$ of \textsf{D} for the tri-gram \textsf{BCD}, we have to map the vocabulary ID of \textsf{D}, i.e., $3$, to $1$. For this task, we search $3$ within the successors of \textsf{C}. As $3$ is found in position $1$, we now know that we have to search for $1$ within the successors of \textsf{BC}.

On the one hand, the context-based remapping will assign smaller IDs as the length of the context rises, on the other hand it will also spend more time at query processing.
Therefore, we have a space/time trade-off that we explore with an extensive experimental analysis in Section~\ref{subsec:indexing:exp}.
The pseudo code for the $\lk$ operations with context-based remapping is illustrated in Figure~\ref{alg:lookup-remapping}. Note that, in comparison with the pseudo code in Figure~\ref{alg:lookup-no-remapping}, the remapping technique uses an array to store the re-mapped IDs (Line 3) and an additional \texttt{for} loop (Lines 7-8).

\parag{Example}
To better understand how the remapping algorithm works, we consider now a small didactic example.
We continue with the example trie from Section~\ref{subsec:indexing:core} and represented in Figure~\ref{fig:trie}.
The darkest blue IDs are the vocabulary IDs and the blue ones are the last token IDs of the tri-grams as assigned by the vocabulary. We now explain how the remapped IDs, represented in light green, are derived by the model using our technique with a context of length $1$. Consider the tri-gram \textsf{BCD}. The default ID of \textsf{D} is $3$. We now rewrite this ID as the position that \textsf{D} takes within the successors of the word preceding it, i.e., the \textsf{C} because we are using a context of length $1$.
As we can see, \textsf{D} appears in position $1$ within the successors of \textsf{C}, therefore its new ID will be $1$. Another example: take \textsf{DBB}. The default ID of \textsf{B} is $1$, but it occurs in position $0$ within the successors of its parent \textsf{B}, therefore its new ID is $0$.
The example in Figure~\ref{fig:trie} illustrates how to map tri-grams using a context of length $1$: this is clearly the only one possible as the first two levels of the trie must be used to retrieve the mapped ID at query time. However, if we have an $n$-gram of order $4$, i.e., $w_1^4$, we can choose to map $w_4$ as the position it takes within the successors of $w_3$ (context of length $1$) or within the successors of $w_2w_3$ (context of length $2$).

\subsection{Hashing}\label{subsec:indexing:hashing}
Since the indexed $n$-gram corpus is static, we obtain a \emph{full} hash utilization by resorting to Minimal Perfect Hash (MPH).
We index all $n$-grams (of the same order $n$) into a separate MPH table, \emph{levels}$[n]$, each with its own MPH function $h_n$.
This introduces a twofold advantage over the linear probing approach used in the literature~\cite{Heafield11,PK11}: use a hash table of size \emph{equal} to the exact number of grams per order (no extra space allocation is required) and avoid the linear probing search phase by requiring one single access to the required hash location.

We use the publicly available implementation of MPH as described in~\cite{BBOVV14} and available at \url{https://github.com/ot/emphf}.
This implementation requires only $2.61$ bits per key on average.

At the hash location for an $n$-gram we store: its $8$-byte hash key as to have a false positive probability of $2^{-64}$ ($4$-byte hash keys are possible as well) and the index of the frequency count in the unique-value array $C[n]$ that keeps all distinct frequency counts for order $n$.
%As already motivated, these unique-value arrays, one for each different order of $n$, are negligibly small compared to the number of $n$-grams themselves and act as a direct map from the position of the count to its value.
Although these unique values could be sorted and compressed, we do not perform any space optimization as these are too few to yield any improvement but we store them uncompressed and byte-aligned, in order to favour lookup speed.
We also use this hash approach to implement the vocabulary of the previously introduced trie data structure.

\parag{Lookup}
Given the $n$-gram $w_1^n$ we compute the position $p = h_n(w_1^n)$ in the relevant table \emph{levels}$[n]$, then we access the count index $i$ stored at position $p$ and finally retrieve the count value $C[n][i]$.

\subsection{Experiments}\label{subsec:indexing:exp}
In this subsection, we first present experiments to validate the effectiveness of our compressed data structures in relation to the corresponding query processing speed; then we compare our proposals against several solutions available in the state-of-the-art.

\parag{Datasets}
We performed our experiments on the following standard datasets.

\begin{itemize}[leftmargin=*]
\item $\euro$ consists in all unpruned $n$-grams extracted from the English Europarl parallel corpus~\cite{europarl}, available at:
\url{http://www.statmt.org/europarl}.

\item $\yahoo$ is a collection of English $n$-grams with minimum frequency count equal to $2$, extracted from a corpus of 14.6 million documents crawled from more than $\num{12 000}$ sites during 2006~\cite{YahooV2}.
The dataset is available at:
\url{http://webscope.sandbox.yahoo.com/catalog.php?datatype=l}.

%This dataset contains n-grams (contiguous sets of words of size n), n = 1 to 5, extracted from a corpus of 14.6 million documents (126 million unique sentences, 3.4 billion running words) crawled from over 12000 news-oriented sites. The documents were published on these sites between February 2006 and December 2006. The dataset does not contain the documents themselves, but only the n-grams that occur at least twice. It provides statistics such as frequency of occurrence, number and entropy of different left (right) single-token contexts of each n-gram. This dataset may be used by researchers to build statistical language models for speech or handwriting recognition or machine translation. There are 3 files in this dataset. They are 3.5 Gbyte, 4.3 Gbyte and 4.4 Gbyte.

\item $\google$ is the latest English version of \textsf{Web1T}~\cite{Web1T}, whose $n$-grams have a minimum frequency count of $40$.
This collection roughly corresponds to $6\%$ of the books ever published.
The dataset is available at:
\url{http://storage.googleapis.com/books/ngrams/books/datasetsv2.html}.
\end{itemize}

Each dataset comprises all $n$-grams for $1 \leq n \leq N = 5$ and associated frequency counts.
Table~\ref{tab:indexing:datasets_stats} shows the basic statistics of the datasets.
These standard datasets are also suitable to test our data structures on different corpora sizes: starting from the left of the table each dataset has roughly $10\times$ the number of $n$-grams of the previous one.

\begin{table}[t]
  \centering
    \scalebox{0.9}{
    \begin{tabular}{c rr rr rr}
        \toprule
        \multirow{2}{*}{$n$} &
\multicolumn{2}{c@{}@{}@{}@{}@{}}{\euro}  &
\multicolumn{2}{c@{}@{}@{}@{}@{}}{\yahoo} &
\multicolumn{2}{c@{}@{}@{}@{}@{}}{\google} \\

\cmidrule(lr){2-3}
\cmidrule(lr){4-5}
\cmidrule(lr){6-7}

& $n$-grams & counts
& $n$-grams & counts
& $n$-grams & counts
\\

\midrule

1 & \num{304579}   & \num{4518}
  & \num{3475482}  & \num{23785}
  & \num{24357349} & \num{246490} \\

2 & \num{5192260}   & \num{4663}
  & \num{53844927}  & \num{31711}
  & \num{665752080} & \num{722966} \\

3 & \num{18908249}   & \num{2975}
  & \num{187639522}  & \num{19856}
  & \num{7384478110} & \num{683653} \\

4 & \num{33862651}   & \num{1744}
  & \num{287562409}  & \num{10761}
  & \num{1642783634} & \num{133491} \\

5 & \num{43160518}   & \num{1032}
  & \num{295701337}  & \num{6167}
  & \num{1413870914} & \num{104025} \\

\midrule

total
	    & \num{101428257}     & \num{7147}
	    & \num{828223677}     & \num{45285}
	    & \num{11131242087} & \num{1073473} \\

%\midrule
%\textsf{gzip} &  6.98 &
%                    &  6.45 &
%                    &  6.20 & \\

%6.9763
%6.4497
%6.2035

        \bottomrule
    \end{tabular}
    }
\caption{Number of $n$-grams and distinct frequency counts for the datasets used in the experiments.
%We also report the average bytes per gram achieved by \textsf{gzip} as a useful reference point for comparison.
}
\vspace{-0.3cm}
\label{tab:indexing:datasets_stats}
\end{table}

\parag{Compared indexes}
We compare the performance of our data structures against the following software packages that use the approaches discussed in Section~\ref{subsec:indexing:related}:
$\berk$ by~\citet{PK11} (\textsf{Java} code at \url{https://github.com/adampauls/berkeleylm});
$\expg$ by~\citet{Watanabe09} (\textsf{C++} code at \url{https://github.com/tarowatanabe/expgram});
$\ken$ by~\citet{Heafield11} (\textsf{C++} code at \url{http://kheafield.com/code/kenlm});
$\mari$ by~\citet{Yata11} (\textsf{C++} code at \url{https://github.com/s-yata/marisa-trie});
$\rand$ by~\citet{TO07} (\textsf{C++} code at \url{https://sourceforge.net/projects/randlm}).

%\begin{itemize}[leftmargin=*]
%\item $\berk$ implements two trie data structures based on sorted arrays and hash tables to represent the nodes of the trie~\cite{PK11}.
%The code is written in \textsf{Java} and available at:
%\url{https://github.com/adampauls/berkeleylm}.
%\item $\expg$ makes use of the \textsf{LOUDS} succinct encoding~\cite{Jacobson89} to implicitly represent the trie structure, while the frequency counts are compressed using \textsf{VByte} encoding~\cite{Watanabe09}.
%The code is written in \textsf{C++} and available at:
%\url{https://github.com/tarowatanabe/expgram}.
%\item $\ken$ implements a trie with interpolation search and a hash table with linear probing~\cite{Heafield11}.
%The code is written in \textsf{C++} and available at:
%\url{http://kheafield.com/code/kenlm}.
%\item $\mari$ is a general-purposes string dictionary implementation in which Patricia tries are recursively used to represent the nodes of a Patricia trie~\cite{Yata11}.
%The code is written in \textsf{C++} and available at:
%\url{https://github.com/s-yata/marisa-trie}.
%\item $\rand$ employs Bloom filters with lossy quantization of frequency counts to attain to low memory footprint~\cite{TO07}.
%The code is written in \textsf{C++} and available at:
%\url{https://sourceforge.net/projects/randlm}.
%\end{itemize}

\parag{Experimental setting and methodology}
All experiments have been performed on a machine with $16$ Intel \textsf{Xeon E5-2630 v3} cores ($32$ threads) clocked at $2.4$ Ghz, with $193$ GB of RAM, running Linux $3.13.0$, $64$ bits.
%Hardware caches have the following sizes: 32 KB (\textsf{L1}), 256 KB (\textsf{L2}) and 20 MB (\textsf{L3}). Levels \textsf{L1} and \textsf{L2} are private of each core, while \textsf{L3} is shared among all the 8 cores on one socket.
Our implementation is in standard \textsf{C++11} and compiled with \textsf{gcc} 5.4.1 with the highest optimization settings, i.e., with compilation flags \texttt{-O3} and \texttt{-march=native}.
To ensure a fair comparison with the other competitors, we used the same compiler and optimization flags for all the \textsf{C++} implementations.

Template specialization has been preferred over inheritance to avoid the virtual method call overhead, which can be disruptive for the very fine-grained operations we consider.
Except for the instructions to count the number of bits set in a word (popcount), and to find the position of the least significant bit (number of trailing zeroes), no special processor feature was used. In particular, we did not add any SIMD (Single Instruction Multiple Data) instruction to our code.

The data structures were saved to disk after construction, and loaded into main memory to be queried.
For the scanning of input files we used the \texttt{posix\_madvice} system, called with the parameter \texttt{POSIX\_MADV\_SEQUENTIAL} to instruct the kernel to optimize the sequential access to the mapped memory region.
The implementation of our data structures, as well as the utilities to prepare the datasets for indexing and unit tests, is freely available at: \url{https://github.com/jermp/tongrams}.

To test the speed of $\lk$ queries, we use a query set consisting of $5$ million $n$-grams for $\yahoo$ and $\google$ and of $0.5$ million for $\euro$, drawn at random from the entire datasets.
In order to smooth the effect of fluctuations during measurements, we repeat each experiment five times and consider the mean.
The query algorithms were run on a single core.

\subsubsection{Elias-Fano tries}\label{subsec:indexing:ef-trie-exp}
In this subsubsection we test the efficiency of the trie data structure described in Section~\ref{subsec:indexing:ef-trie}.
%As already done for the description in Section~\ref{subsec:indexing:core}, we dedicate one paragraph to the validation of each of the main building components of the trie, as well as to the introduced performance optimizations.

\parag{Gram-ID sequences}
Table~\ref{tab:bpg} shows the average number of bytes per gram including the cost of pointers, and lookup speed per query.
The first two rows refers to the trie data structure described in Section \ref{subsec:indexing:core}, when the sorted arrays are encoded with Elias-Fano ($\EF$) and partitioned Elias-Fano ($\PEF$)~\cite{OV14}.
Subsequent rows indicate the space gains obtained by applying the context-based remapping strategy using $\EF$ and $\PEF$ for contexts of lengths respectively $1$ and $2$.
For $\google$ we use a context of length $1$, as the tri-grams alone roughly constitute $66\%$ of the whole the dataset, thus it would make little sense to optimize only the space of $4$- and $5$-grams that take (together) $27\%$ of the dataset.

As expected, partitioning the gram sequences using $\PEF$ yields a better space occupancy.
Though the paper by Ottaviano and Venturini~\cite{OV14} describes a dynamic programming algorithm to find the partitioning that minimizes the space occupancy of a monotone sequence, we instead adopt a \emph{uniform} partitioning strategy. Partitioning the sequence uniformly has several advantages over variable-length partitions for our setting.
As we have seen in Section~\ref{subsec:indexing:core}, trie searches are carried out by performing a preliminary random access to the endpoints of the range pointed to by a pointer pair.
Then a search in the range follows to determine the position of the gram-ID.
Partitioning the sequence by variable-length blocks introduces an additional search over the sequence of partition endpoints to determine the proper block in which the search must continue. While this preliminary search only introduces a minor overhead in query processing for inverted index queries~\cite{OV14} (as it has to be performed once and successive accesses are only directed to \emph{forward} positions of the sequence), it is instead the major bottleneck when random access operations are very frequent as in our case.
By resorting on uniform partitions, we eliminate this first search \emph{and} the cost of representation for the variable-length sizes.
To speed up queries even further, we also keep the upper bounds of the blocks uncompressed and bit-aligned.

\begin{table}[t]
\scalebox{0.9}{
    \begin{tabular}{l@{\hspace{2pt}}
                          l@{\hspace{5pt}}
                          l@{\hspace{5pt}}
                          l@{\hspace{15pt}}
                          r@{\hspace{2pt}}
                          r@{\hspace{6pt}}
                          r@{\hspace{2pt}}
                          r@{\hspace{0pt}}
                          r@{\hspace{15pt}}
                          r@{\hspace{2pt}}
                          r@{\hspace{6pt}}
                          r@{\hspace{2pt}}
                          r@{\hspace{0pt}}
                          r@{\hspace{15pt}}
                          r@{\hspace{2pt}}
                          r@{\hspace{6pt}}
                          r@{\hspace{2pt}}
                          r@{\hspace{0pt}}}
    \toprule
    
            & & &
            & \multicolumn{4}{c@{}@{}@{}@{}}{\euro}
            &
            & \multicolumn{4}{c@{}@{}@{}@{}}{\yahoo}
            &
            & \multicolumn{4}{c@{}@{}@{}@{}}{\google} \\
            
            \cmidrule(lr){5-8}
            \cmidrule(lr){10-13}
            \cmidrule(lr){15-18}

            & & &
            & \multicolumn{2}{@{}l@{}}{$\bpg$}
            & \multicolumn{2}{@{}l@{}}{$\mpq$}
            &
            & \multicolumn{2}{@{}l@{}}{$\bpg$}
            & \multicolumn{2}{@{}l@{}}{$\mpq$}
            &
            & \multicolumn{2}{@{}l@{}}{$\bpg$}
            & \multicolumn{2}{@{}l@{}}{$\mpq$} \\

            \cmidrule(lr){4-18}
            
            & &
            & \EF
            & $1.97$ &
            & $1.28$ &
            &
            & $2.17$ &
            & $1.60$ &
            &
            & $2.13$ &
            & $2.09$ & \\

            & &
            & \PEF
            & $1.87$ & {\color{DarkGray}{$(-5\%)$}}
            & $1.35$ & {\color{DarkGray}{$(+6\%)$}}
            &
            & $1.91$ & {\color{DarkGray}{$(-12\%)$}}
            & $1.73$ & {\color{DarkGray}{$(+8\%)$}}
            &
            & $1.52$ & {\color{DarkGray}{$(-29\%)$}}
            & $1.91$ & {\color{DarkGray}{$(-9\%)$}} \\
            
            \cmidrule(lr){4-18}

            \multirow{4}{*}{\rotatebox[origin=c]{90}{{\fontsize{2.2mm}{2.2mm}\selectfont \sf CONTEXT-BASED}}}
            &
            \multirow{4}{*}{\rotatebox[origin=c]{90}{{\fontsize{2.5mm}{2.5mm}\selectfont \sf ID REMAPPING}}}

            & \multirow{2}{*}{\rotatebox[origin=c]{90}{{\fontsize{2.5mm}{2.5mm}\selectfont $k = 1$}}}
            & \EF
            & $1.67$ & {\color{DarkGray}{$(-15\%)$}}
            & $1.58$ & {\color{DarkGray}{$(+24\%)$}}
            &
            & $1.89$ & {\color{DarkGray}{$(-13\%)$}}
            & $2.05$ & {\color{DarkGray}{$(+28\%)$}}
            &
            & $1.91$ & {\color{DarkGray}{$(-10\%)$}}
            & $3.03$ & {\color{DarkGray}{$(+45\%)$}} \\
            
            & &
            & \PEF
            & $1.53$ & {\color{DarkGray}{$(-22\%)$}}
            & $1.61$ & {\color{DarkGray}{$(+26\%)$}}
            &
            & $1.63$ & {\color{DarkGray}{$(-25\%)$}}
            & $2.16$ & {\color{DarkGray}{$(+35\%)$}}
            &
            & $1.31$ & {\color{DarkGray}{$(-39\%)$}}
            & $2.30$ & {\color{DarkGray}{$(+10\%)$}} \\

            \cmidrule(lr){4-18}

            & & \multirow{2}{*}{\rotatebox[origin=c]{90}{{\fontsize{2.3mm}{2.3mm}\selectfont $k = 2$}}}
            & \EF
            & $1.46$ & {\color{DarkGray}{$(-26\%)$}}
            & $1.60$ & {\color{DarkGray}{$(+25\%)$}}
            &
            & $1.68$ & {\color{DarkGray}{$(-22\%)$}}
            & $2.08$ & {\color{DarkGray}{$(+30\%)$}}
            & & --- & & --- & \\

            & &
            & \PEF
            & $1.28$ & {\color{DarkGray}{$(-35\%)$}}
            & $1.64$ & {\color{DarkGray}{$(+28\%)$}}
            &
            & $1.38$ & {\color{DarkGray}{$(-36\%)$}}
            & $2.15$ & {\color{DarkGray}{$(+35\%)$}}
            & & --- & & --- & \\

    \bottomrule
    \end{tabular}
}
\caption{Average bytes per gram ($\bpg$) and average $\lk$ time per query in micro seconds ($\mpq$). The $\bpg$ cost also includes the space of representation for the pointer sequences.}
\label{tab:bpg}
\vspace{-0.6cm}
\end{table}

\begin{figure}
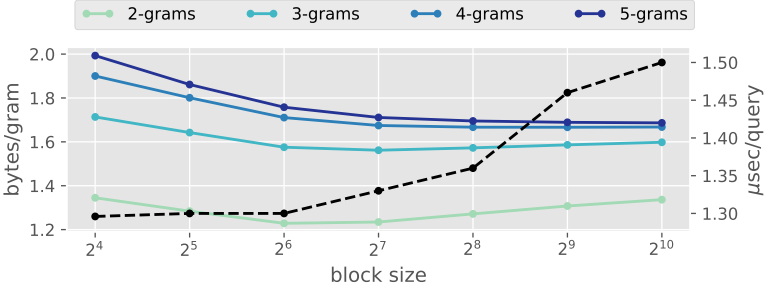

    \includegraphics[scale=0.62]{{{plots/europarl.blocks}}}
    \caption{Bytes per gram (left vertical axis) and $\mu$s per query (right vertical axis, black dashed line) by varying block size in \textsf{PEF uniform} on the gram-ID sequences of $\euro$.}    
     \label{fig:europarl_blocks}
\end{figure}

As the problem of deciding the optimal block size is posed, Figure~\ref{fig:europarl_blocks} shows the space/time trade-off obtained by varying the block size on the gram-ID sequences.
The plots for $\yahoo$ and $\google$ datasets exhibit the same shape, therefore we show the one for the $\euro$ dataset.
The dashed black line illustrates how the average $\lk$ time varies when \emph{all} the gram-ID sequences are partitioned using the same block size.
The figure suggests to use partitions of $64$ integers for bigram sequences, and of $128$ for all other orders, i.e., for $N \geq 3$, given that the space usage remains low without increasing much the query processing speed.
With this choice of block sizes, the increase in space consumption with respect to the optimal partitioning is small and equal to $3.32\%$ for $\euro$; $5.29\%$ for $\yahoo$ and $7.33\%$ for $\google$.

Shrinking the size of blocks speeds up searches over plain Elias-Fano because a successor query has to be resolved over an interval potentially much smaller than a range length. This behaviour is clearly highlighted by the shape of the black dashed line of Figure~\ref{fig:europarl_blocks}.
However, excessively reducing the block size may ruin the advantage in space reduction.
Therefore it is convenient to use small block sizes for the most traversed sequences, e.g., the bigram sequences, that indeed must be searched several times during the query-mapping phase when the context-based remapping is adopted.
In conclusion, as we can see by the second row of Table~\ref{tab:bpg},
there is \emph{no} practical difference between the query processing speed of $\EF$ and $\PEF$:
this latter sequence organization brings a negligible overhead in query processing speed (less than $8\%$ on $\euro$ and $\yahoo$), while maintaining a noticeable space reduction (up to $29\%$ on $\google$).

\parag{Context-based identifier remapping}
Concerning the effectiveness of the context-based remapping, we can see from Table~\ref{tab:bpg} that remapping the gram IDs with a context of length $k=1$ is already able of reducing the space of the sequences by ${\approx}13\%$ on average when sequences are encoded with Elias-Fano, with respect to the $\EF$ cost. If we consider a context of length $k=2$ we \emph{double} the gain, allowing for more than $28\%$ of space reduction \emph{without} affecting the lookup time with respect to the case $k=1$.
The lookup speeds for $k=1$ and $k=2$ are pretty much the same because the number of successors for a bigram is very small on average (as already noted with the numbers shown in Figure~\ref{tab:stats}), therefore a search among very few successors (case for $k=2$) results in an almost negligible time overhead with respect to the case of one single search (case for $k=1$).
%Note that the fact that the number of successors per $n$-gram reduces dramatically as $n$ increases also explains why the context-based remapping is so effective, as already noted by the numbers shown in Figure~\ref{tab:stats}.

As a first conclusion, when space efficiency is the main concern, it is always convenient to apply the remapping strategy with a context of length $2$.
The gain of the strategy is even more evident with $\PEF$: this is no surprise as the encoder can better exploit the reduced IDs by encoding all the integers belonging to a block with a universe relative to the block and not to the whole sequence.
This results in a space reduction of more than $36\%$ on average and up to $39\%$ on $\google$.

Regarding the query processing speed, as explained in Section~\ref{subsec:indexing:remapping}, the remapping strategy comes with a penalty at query time as we have to map an ID before it can be searched in the proper gram sequence.
On average, by looking at Table~\ref{tab:bpg}, we found that $30\%$ more time is spent with respect to the Elias-Fano baseline. Notice that $\PEF$ does \emph{not} introduce any time degradation with respect to $\EF$ with context-based remapping: it is actually faster on $\google$.

\begin{table}
    \centering
    \scalebox{0.9}{
    \begin{tabular}{
                    l@{\hspace{15pt}}
                    r@{\hspace{2pt}}
                    r@{\hspace{15pt}}
                    r@{\hspace{2pt}}
                    r@{\hspace{15pt}}
                    r@{\hspace{2pt}}
                    r@{\hspace{2pt}}
                    }
    \toprule
    
			& \multicolumn{2}{c@{}@{}@{}@{}}{\euro}
			& \multicolumn{2}{c@{}@{}@{}@{}}{\yahoo}
			& \multicolumn{2}{c@{}@{}@{}@{}}{\google} \\

			\cmidrule(lr){1-7}

			\textsf{Variable-len. codewords}
			& $0.36$ &
			& $\mathbf{0.47}$ &
			& $1.46$ &
			\\

			\textsf{Prefix sums} + $\EF$
			& $0.35$ & {\color{DarkGray}{$(-2\%)$}}
			& $0.62$ & {\color{DarkGray}{$(+33\%)$}}
			& $1.59$ & {\color{DarkGray}{$(+9\%)$}}
			\\

			\textsf{Prefix sums} + $\PEF$
			& $\mathbf{0.30}$ & {\color{DarkGray}{$(-17\%)$}}
			& $0.51$ & {\color{DarkGray}{$(+9\%)$}}
			& $\mathbf{1.30}$ & {\color{DarkGray}{$(-11\%)$}}
			\\

			\cmidrule(lr){1-7}

			\textsf{Variable-len. block-coding}
			& $0.76$ & {\color{DarkGray}{$(+156\%)$}}
			& $0.79$ & {\color{DarkGray}{$(+56\%)$}}
			& $1.32$ & {\color{DarkGray}{$(+1\%)$}}
			\\

			\textsf{Packed}
			& $1.63$ & {\color{DarkGray}{$(+445\%)$}}
			& $2.00$ & {\color{DarkGray}{$(+294\%)$}}
			& $2.63$ & {\color{DarkGray}{$(+102\%)$}}
			\\

			\cmidrule(lr){1-7}

			\textsf{VByte}
			& $3.21$ & {\color{DarkGray}{$(+975\%)$}}
			& $3.32$ & {\color{DarkGray}{$(+555\%)$}}
			& --- &
			\\

    \bottomrule
    \end{tabular}
    }
\caption{Average bytes per count for different techniques.}
\label{tab:bpc}
\vspace{-0.6cm}
\end{table}

\parag{Frequency counts}
For the representation of frequency counts we compare three different encoding schemes: the first one refers to the strategy described in Section~\ref{subsec:indexing:core} that assigns variable-length codewords to the ranks of the counts and keeps track of codewords length using a binary vector (\textsf{Variable-len. codewords});
the other two schemes transform the sequence of count ranks into a non-decreasing sequence by taking its prefix sums and then applies $\EF$ or $\PEF$ (\textsf{Prefix sums + $\EF$/$\PEF$}).

Table~\ref{tab:bpc} shows the average number of bytes per count for these different strategies.
The reported space also includes the space for the storage of the arrays containing the distinct counts for each order of $N$. As already pointed out, these take a negligible amount of space because the distribution of frequency counts is highly repetitive (see Table~\ref{tab:indexing:datasets_stats}).
The percentages of \textsf{Prefix sums + $\EF$/$\PEF$} are done with respect to the first row of the table, i.e., \textsf{Variable-len. codewords}.

The time for retrieving a count was pretty much the same for all the three techniques.
Prefix-summing the sequence and applying $\EF$ does not bring any advantage over the codeword assignment technique because its space is practically the same on $\euro$ but it is actually larger on both $\yahoo$ (by up to $32\%$) and $\google$.
These two reasons together place the codeword assignment technique in net advantage over $\EF$.
$\PEF$, instead, offers a better space occupancy of more than $16\%$ on $\euro$ and $10\%$ on $\google$. Therefore, in the following we assume this representation for frequency counts, except for $\yahoo$, where we adopt \textsf{Variable-len. codewords}.

We also report the space occupancy for the counts representation of $\berk$ and $\expg$ which, differently from all other competitors, can also be used to index frequency counts.
$\berk$ \textsf{COMPRESSED} variant uses the \textsf{Variable-len. block-coding} mechanism explained in Section~\ref{subsec:indexing:related} to compress count ranks, whereas the \textsf{HASH} variant stores bit-packed count ranks, referred to as \textsf{Packed} in the table, using the minimum number of bits necessary for their representation (see Table~\ref{tab:indexing:datasets_stats}).
$\expg$, instead, does not store count ranks but directly compress the counts themselves using Variable-Byte encoding (\textsf{VByte}) with an additional binary vector as to be able of randomly accessing the counts sequence.
The available RAM of our test machine ($193$ GB) was not sufficient to successfully build $\expg$ on $\google$. The same holds for $\ken$ and $\mari$, as we are going to see next. Therefore, we report its space for $\euro$ and $\yahoo$.

We first observe that rank-encoding schemes are far more advantageous than compressing the counts themselves, as done by $\expg$. Moreover, none of these techniques beats the three techniques we previously introduced, except for the $\berk$ \textsf{COMPRESSED} variant which is ${\approx} 10\%$ smaller on $\google$ with respect to \textsf{Variable-len. codewords}.
However, note that this gap is completely bridged as soon as we adopt the combination \textsf{Prefix sums + $\PEF$}.

\begin{figure}
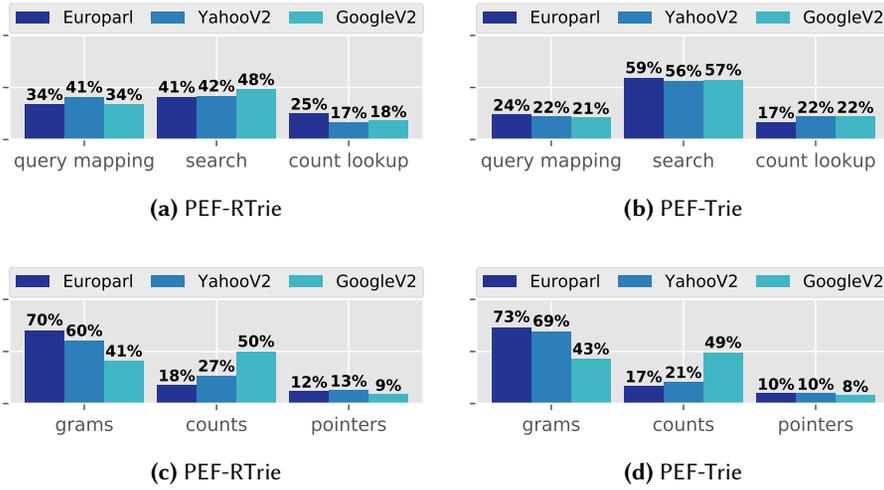

    \vspace{-0.5cm}
    \centering
    \subfloat[$\spaceopt$]{
    \includegraphics[scale=0.7]{{{figures/time_breakdowns.space_opt.trie}}}
    \label{fig:time_spaceopt}}
    \subfloat[$\timeopt$]{
    \includegraphics[scale=0.7]{{{figures/time_breakdowns.time_opt.trie}}}
    \label{fig:time_timeopt}}
 
    \subfloat[$\spaceopt$]{
    \includegraphics[scale=0.7]{{{figures/size_breakdowns.space_opt.trie}}}
    \label{fig:size_spaceopt}}
    \subfloat[$\timeopt$]{
    \includegraphics[scale=0.7]{{{figures/size_breakdowns.time_opt.trie}}}
    \label{fig:size_timeopt}}
    
    \caption{Trie data structures timing (\textsf{a-b}) and size (\textsf{c-d}) breakdowns in percentage on the tested datasets.
    For the timing breakdowns we distinguish the three phases of query mapping, ID-search and final count lookup. For the space breakdowns we distinguish, instead, the contribution of gram-ID, count and pointer sequences.}
\end{figure}

\parag{Time and space breakdowns}
Now we use the analysis done so far to fix two different trie data structures that, respectively, privilege space efficiency and query time: we call them $\spaceopt$ (the \textsf{R} stands for \emph{remapped}) and $\timeopt$.
For the $\spaceopt$ variant we use $\PEF$ for representing the gram-ID sequences; \textsf{Prefix sums + $\PEF$} for the counts on $\euro$ and $\google$ but \textsf{Variable-len. codewords} for $\yahoo$.
We also use the maximum applicable context length for the context-based remapping technique, i.e., $2$ for $\euro$ and $\yahoo$; $1$ for $\google$.
For the $\timeopt$ variant we choose a data structure using $\PEF$ for representing gram-ID sequences and \textsf{Variable-len. codewords} for the counts, \emph{without} remapping.

The corresponding size breakdowns are shown in Figure~\ref{fig:size_spaceopt} and Figure~\ref{fig:size_timeopt} respectively.
Pointer sequences take very little space for both data structures (approximately $10.3\%$), while most of the difference lies, not surprisingly, in the space of the gram-ID sequences (roughly $70\%$ for $\euro$ and $\yahoo$; $40\%$ for $\google$).
Instead, the timing breakdowns in Figure~\ref{fig:time_spaceopt} and Figure~\ref{fig:time_timeopt} clearly highlight how the context-based remapping technique \emph{rises} the time we spend in the query-mapping phase, during which the IDs are mapped to their reduced IDs. In such case, the two phases of query mapping (given by vocabulary lookups plus context-based remapping) and search are almost the same, while in the $\timeopt$ the search phase dominates.

%\begin{table}
%   \centering
%   \begin{tabular}{
%                         l@{\hspace{5pt}}
%                         r@{\hspace{3pt}}
%                         r@{\hspace{3pt}}
%                         r@{\hspace{3pt}}
%                         r@{\hspace{3pt}}
%                         r@{\hspace{3pt}}
%                         r@{\hspace{3pt}}
%                         r@{\hspace{3pt}}
%                         r@{\hspace{3pt}}
%                         r@{\hspace{3pt}}
%                         }
%   \toprule
%   \input{tables/hash_bpg_bpc_lookup}
%   \bottomrule
%   \end{tabular}
%\caption{Average bytes per gram (\textsf{bpg}); average bytes per count (\textsf{bpc}) and average $\lk$ time in micro seconds for a $\MPH$ data structure using $8$-bytes hash keys.}
%\label{tab:hash_bpg_bpc_lookup}
%\end{table}

\subsubsection{Hashing}\label{subsec:indexing:hashing-exp}
%Table \ref{tab:hash_bpg_bpc_lookup} shows the results for the minimal perfect hash ($\MPH$) data structure.
We build our $\MPH$ tables using $8$-byte hash keys, as to yield a false positive rate of $2^{-64}$. For each different value of $n$ we store the distinct count values in an array, uncompressed and byte-aligned using $4$ bytes per distinct count on $\euro$ and $\yahoo$; $8$ bytes on $\google$.

For all the three datasets, the number of bytes per gram, including also the cost of the hash function itself ($0.33$ bytes per gram) is $8.33$. The number of bytes per count is given by the sum of the cost for the ranks and the distinct counts themselves and is equal to $1.41$, $1.74$ and $2.43$ for $\euro$, $\yahoo$ and $\google$ respectively.
Not surprisingly, the majority of space is taken by the hash keys: clients willing to reduce this memory impact can use $4$-byte hash keys instead, at the price of a higher false positive rate ($2^{-32}$).
Therefore, it is worth observing that spending additional effort in trying to lower the space occupancy of the counts only results in poor improvements as we pay for the high cost of the hash keys.

The constant-time access capability of hashing makes gram lookup extremely fast, by requiring on average $1/3$ of a micro second per lookup (exact numbers are reported in Table~\ref{tab:overall_comparison}). In particular, all the time is spent in computing the hash function itself and access the relative table location: the final count lookup is completely negligible.

\begin{table}
    \centering
    \scalebox{0.9}{
    \begin{tabular}{
        l@{\hspace{10pt}}
        r@{\hspace{2pt}}
        r@{\hspace{4pt}}
        r@{\hspace{2pt}}
        r@{\hspace{0pt}}
        r@{\hspace{10pt}}
        r@{\hspace{2pt}}
        r@{\hspace{4pt}}
        r@{\hspace{2pt}}
        r@{\hspace{0pt}}
        r@{\hspace{10pt}}
        r@{\hspace{2pt}}
        r@{\hspace{4pt}}
        r@{\hspace{2pt}}
        r@{\hspace{0pt}}
    }
    \toprule
    
			& 	\multicolumn{4}{c@{}@{}@{}@{}@{}@{}}{\euro}
			&
			& 	\multicolumn{4}{c@{}@{}@{}@{}@{}@{}}{\yahoo}
			&
			& 	\multicolumn{4}{c@{}@{}@{}@{}@{}@{}}{\google} \\

			\cmidrule(lr){2-5}
			\cmidrule(lr){7-10}
			\cmidrule(lr){12-15}

			&	\multicolumn{2}{@{}l@{}}{$\msp$ $\bpg$}
			&	\multicolumn{2}{@{}l@{}}{$\mpq$}
			&
			&	\multicolumn{2}{@{}l@{}}{$\msp$ $\bpg$}
			&	\multicolumn{2}{@{}l@{}}{$\mpq$}
			&
			&	\multicolumn{2}{@{}l@{}}{$\bpg$}
			&	\multicolumn{2}{@{}l@{}}{$\mpq$} \\

			\cmidrule(lr){1-15}

			\timeopt
						& $1.87$ &
						& $1.35$ &
						&
						& $1.91$ &
						& $1.73$ &
						&
						& $1.52$ &
						& $\mathbf{1.91}$ & \\

			\spaceopt
						& $\mathbf{1.28}$ &
						& $1.64$ &
						&
						& $\mathbf{1.38}$ &
						& $2.15$ &
						&
						& $\mathbf{1.31}$ &
						& $2.30$ & \\

			\cmidrule(lr){1-15}

			{\textsf{Berk. C.}}
						& $1.70$ & {\color{DarkGray}{$(-9\%)$}}
						& $2.83$ & {\color{DarkGray}{$(+109\%)$}}
						&
						& $1.69$ & {\color{DarkGray}{$(-11\%)$}}
						& $3.48$ & {\color{DarkGray}{$(+102\%)$}}
						&
						& $1.45$ & {\color{DarkGray}{$(-5\%)$}}
						& $4.13$ & {\color{DarkGray}{$(+117\%)$}}
						\\

						& & {\color{DarkGray}{$(+33\%)$}}
						& & {\color{DarkGray}{$(+73\%)$}}
						&
						& & {\color{DarkGray}{$(+22\%)$}}
						& & {\color{DarkGray}{$(+62\%)$}}
						&
						& & {\color{DarkGray}{$(+11\%)$}}
						& & {\color{DarkGray}{$(+80\%)$}}
						\\

			{\textsf{Berk. H.3}}
						& $6.70$ & {\color{DarkGray}{$(+259\%)$}}
						& $0.97$ & {\color{DarkGray}{$(-29\%)$}}
						&
						& $7.82$ & {\color{DarkGray}{$(+310\%)$}}
						& $1.13$ & {\color{DarkGray}{$(-34\%)$}}
						&
						& $9.24$ & {\color{DarkGray}{$(+508\%)$}}
						& $2.18$ & {\color{DarkGray}{$(+14\%)$}}
						\\

						& & {\color{DarkGray}{$(+423\%)$}}
						& & {\color{DarkGray}{$(-41\%)$}}
						&
						& & {\color{DarkGray}{$(+465\%)$}}
						& & {\color{DarkGray}{$(-47\%)$}}
						&
						& & {\color{DarkGray}{$(+608\%)$}}
						& & {\color{DarkGray}{$(-5\%)$}}
						\\

			{\textsf{Berk. H.50}}
						& $7.96$ & {\color{DarkGray}{$(+326\%)$}}
						& $\mathbf{0.97}$ & {\color{DarkGray}{$(-29\%)$}}
						&
						& $9.37$ & {\color{DarkGray}{$(+391\%)$}}
						& $\mathbf{0.96}$ & {\color{DarkGray}{$(-44\%)$}}
						& & --- & & --- &
						\\

						  & & {\color{DarkGray}{$(+522\%)$}}
						  & & {\color{DarkGray}{$(-41\%)$}}
						  &
						  & & {\color{DarkGray}{$(+577\%)$}}
						  & & {\color{DarkGray}{$(-55\%)$}}

						& & & & \\

			\expg
						& $2.06$ & {\color{DarkGray}{$(+10\%)$}}
						& $2.80$ & {\color{DarkGray}{$(+107\%)$}}
						&
						& $2.24$ & {\color{DarkGray}{$(+17\%)$}}
						& $9.23$ & {\color{DarkGray}{$(+435\%)$}}
						& & --- & & --- &
						\\

						  & & {\color{DarkGray}{$(+61\%)$}}
						  & & {\color{DarkGray}{$(+71\%)$}}
						  &
						  & & {\color{DarkGray}{$(+62\%)$}}
						  & & {\color{DarkGray}{$(+329\%)$}}
						& & & & \\

			\ken $\msp$ \sf T.
						& $2.99$ & {\color{DarkGray}{$(+60\%)$}}
						& $1.28$ & {\color{DarkGray}{$(-6\%)$}}
						&
						& $3.44$ & {\color{DarkGray}{$(+80\%)$}}
						& $1.94$ & {\color{DarkGray}{$(+12\%)$}}
						& & --- & & --- &
						\\

						  & & {\color{DarkGray}{$(+134\%)$}}
						  & & {\color{DarkGray}{$(-22\%)$}}
						  &
						  & & {\color{DarkGray}{$(+149\%)$}}
						  & & {\color{DarkGray}{$(-10\%)$}}
						& & & & \\

			\mari
						& $3.61$ & {\color{DarkGray}{$(+93\%)$}}
						& $2.06$ & {\color{DarkGray}{$(+52\%)$}}
						&
						& $3.81$ & {\color{DarkGray}{$(+100\%)$}}
						& $3.24$ & {\color{DarkGray}{$(+88\%)$}}
						& & --- & & --- &
						\\

						  & & {\color{DarkGray}{$(+182\%)$}}
						  & & {\color{DarkGray}{$(+26\%)$}}
						  &
						  & & {\color{DarkGray}{$(+175\%)$}}
						  & & {\color{DarkGray}{$(+51\%)$}}
						& & & & \\

			\cmidrule(lr){1-15}

			\rand
						& $1.81$ & {\color{DarkGray}{$(-3\%)$}}
						& $4.39$ & {\color{DarkGray}{$(+224\%)$}}
						&
						& $2.02$ & {\color{DarkGray}{$(+6\%)$}}
						& $5.08$ & {\color{DarkGray}{$(+194\%)$}}
						&
						& $2.60$ & {\color{DarkGray}{$(+71\%)$}}
						& $9.25$ & {\color{DarkGray}{$(+385\%)$}}
						\\

						  & & {\color{DarkGray}{$(+41\%)$}}
						  & & {\color{DarkGray}{$(+168\%)$}}
						&
						  & & {\color{DarkGray}{$(+46\%)$}}
						  & & {\color{DarkGray}{$(+136\%)$}}
						&
						  & & {\color{DarkGray}{$(+99\%)$}}
						  & & {\color{DarkGray}{$(+302\%)$}}
								\\

			\cmidrule(lr){1-15}

			\MPH
						& $\mathbf{8.33}$ &
						& $\mathbf{0.26}$ &
						&
						& $\mathbf{8.33}$ &
						& $\mathbf{0.32}$ &
						&
						& $\mathbf{8.33}$ &
						& $\mathbf{0.37}$ & \\

			\ken $\msp$ \sf P.3
						& $9.40$ & {\color{DarkGray}{$(+13\%)$}}
						& $0.43$ & {\color{DarkGray}{$(+63\%)$}}
						&
						& $9.41$ & {\color{DarkGray}{$(+13\%)$}}
						& $0.38$ & {\color{DarkGray}{$(+20\%)$}}
						& & --- & & --- &
						\\

			\ken $\msp$ \sf P.50
						& $16.91$ & {\color{DarkGray}{$(+103\%)$}}
						& $0.31$ & {\color{DarkGray}{$(+17\%)$}}
						&
						& $16.92$ & {\color{DarkGray}{$(+103\%)$}}
						& $0.34$ & {\color{DarkGray}{$(+8\%)$}}
						& & --- & & --- &
						\\

    \bottomrule
    \end{tabular}
    }
\caption{Average bytes per gram ($\bpg$) and average $\lk$ time per query in micro seconds per query ($\mpq$). For our data structures, i.e., $\timeopt$ and $\spaceopt$, the $\bpg$ cost also includes the space of representation for the pointer sequences.
\textsf{Berk.} is a short-hand name for {\berk}.}
\label{tab:overall_comparison}
\vspace{-0.5cm}
\end{table}

\subsubsection{Overall comparison}\label{subsec:indexing:overall}
We now compare the performance of our selected trie-based solutions, i.e., the $\spaceopt$ and $\timeopt$, as well as our minimal perfect hash approach against the competitors mentioned at the beginning of the experiments.
The results of the comparison are shown in Table~\ref{tab:overall_comparison}, where we report the space taken by the representation of the gram-ID sequences and average $\lk$ time per query in micro seconds. For the trie data structures, the reported space also includes the cost of representation for the pointers.
We compare the space of representation for the $n$-grams excluding their associated information because this varies according to the chosen implementation: for example, $\ken$ can only store probabilities and backoffs, whereas $\berk$ can be used to store either counts or probabilities. For those competitors storing frequency counts, we already discussed their count representation in Section~\ref{subsec:indexing:ef-trie-exp}.
$\expg$, $\ken$ and $\mari$ require too much memory for the building of their data structures on $\google$, therefore we mark as empty their entry in the table for this dataset.

Except for the last two rows of the table in which we compare the performance of our $\MPH$ table against $\ken$ probing (\textsf{P.}), we write for each competitor two percentages indicating its score against our selected trie data structures $\timeopt$ and $\spaceopt$, respectively.

Let us now examine each row, one by one.
In the following discussion, unless explicitly stated, the numbers cited as percentages refer to \emph{average} values over the different datasets.

\return
$\berk$ (\textsf{Berk.}) \textsf{COMPRESSED} (\textsf{C.}) variant results $21\%$ larger than our $\spaceopt$ implementation and slower by more than $70\%$. It gains, instead, an advantage of roughly $9\%$ over our $\timeopt$ data structure, but it is also more than $2$ times slower.
The \textsf{HASH} variant uses hash tables with linear probing to represent the nodes of the trie.
Therefore, we test it with a small extra space factor of $3\%$ for table allocation (\textsf{H.3}) and with $50\%$ (\textsf{H.50}), which is also used as the default value in the implementation, as to obtain different time/space trade-offs.
Clearly the space occupancy of both hash variants do not compete with the ones of our proposals as these are from $3$ to $7$ times larger, but the $\Oh(1)$-lookup capabilities of hashing makes it faster than a sorted array trie implementation: while this is no surprise, notice that our $\timeopt$ data structure is anyway competitive as it is actually faster on $\google$.

$\expg$ is $13.5\%$ larger than $\timeopt$ and also $2$ and $5$ times slower on $\euro$ and $\yahoo$ respectively. Our $\spaceopt$ data structure retains an advantage in space of $60\%$ and it is still significantly faster: of about $70\%$ on $\euro$ and $4.3$ times on $\yahoo$.

$\ken$ is the fastest trie language model implementation in the literature. As we can see, our $\timeopt$ variant retains $70\%$ of its space with a negligible penalty at query time. Compared to the $\spaceopt$ data structure, $\ken$'s trie is slightly faster, i.e., $15\%$, but also $2.3$ and $2.5$ times larger on $\euro$ and $\yahoo$ respectively.

We also tested the performance of $\mari$ even though it is not a trie optimized for language models as to understand how our data structures compare against a general-purpose string dictionary implementation. We outperform $\mari$ in both space and time: compared to $\spaceopt$, it is $2.7$ times larger and $38\%$ slower; with respect to $\timeopt$ it is more than $90\%$ larger and $70\%$ slower.

$\rand$ is designed for small memory footprint and returns approximated frequency counts when queried. We build its data structures using the default setting recommended in the documentation: $8$ bits for frequency count quantization and $8$ bits per value as to yield a false positive rate of $\frac{1}{256}$.
While being from $2.3$ to $5$ times slower than our exact and lossless approach, it is quite compact because the quantized frequency counts are recomputed on the fly using the procedure described in Section~\ref{subsec:indexing:related}.
However, its space is even larger than the one of our $n$-gram representation by $61\%$. It is also larger than the whole space of our $\spaceopt$ data structure. With respect to the whole space of $\timeopt$, it retains instead an advantage of $15.6\%$.
This space advantage is, however, compensated by a loss in precision and a much higher query time (up to $5$ times slower on $\google$).

\return
The last two rows of Table~\ref{tab:overall_comparison} show the performance of our $\MPH$ table with respect to $\ken$ \textsf{PROBING}.
As similarly done for $\berk$ \textsf{H.}, we also test the \textsf{PROBING} data structure with $3\%$ (\textsf{P.3}) and $50\%$ (\textsf{P.50}) extra space allocation factor for the tables.
While being larger as expected, the $\ken$ implementation makes use of expensive hash key recombinations that yields a slower random access capability with respect to our minimal perfect hashing approach.

%\return
%We finally compare the \emph{total} space occupancy, as given by the sum of the space of gram-ID sequences, frequency counts and pointers, of our trie data structures against the \textsf{gzip} baseline reported in Table~\ref{tab:indexing:datasets_stats}.
%The total average bytes per represented $n$-gram for $\timeopt$ are $2.17$, $2.38$ and $2.82$ on the three datasets $\euro$, $\yahoo$ and $\google$ respectively. Table~\ref{tab:indexing:datasets_stats} shows that \textsf{gzip} takes, instead, $6.98$, $6.45$ and $6.2$ bytes per gram.
%This means that our $\timeopt$ is $3.2\times$, $2.7\times$ and $2.2\times$ smaller than \textsf{gzip} and it does also support efficient search of individual $n$-grams.
%Finally, our $\spaceopt$ is $4.4\times$, $3.5\times$, $2.4\times$ smaller.

\begin{table}
    \centering
    \scalebox{0.9}{
    \begin{tabular}{
        l@{\hspace{10pt}}
        r@{\hspace{2pt}}
        r@{\hspace{6pt}}
        r@{\hspace{2pt}}
        r@{\hspace{0pt}}
        r@{\hspace{10pt}}
        r@{\hspace{2pt}}
        r@{\hspace{6pt}}
        r@{\hspace{2pt}}
        r@{\hspace{0pt}}
    }
    \toprule
    		& \multicolumn{4}{c@{}@{}@{}@{}}{\euro}
		&
		& \multicolumn{4}{c@{}@{}@{}@{}}{\yahoo} \\

		\cmidrule(lr){2-5}
		\cmidrule(lr){7-10}

		& \multicolumn{2}{@{}l@{}}{$\msp$ $\bpg$}
		& \multicolumn{2}{@{}l@{}}{$\mpq$}
		&
		& \multicolumn{2}{@{}l@{}}{$\msp$ $\bpg$}
		& \multicolumn{2}{@{}l@{}}{$\msp$ $\mpq$}
		\\

		\cmidrule(lr){1-10}

		\timeopt
		& $3.48$ &
		& $0.25$ &
		&
		& $3.64$ &
		& $0.38$ & \\

		\spaceopt
		& $\mathbf{2.91}$ &
		& $0.28$ &
		&
		& $\mathbf{3.06}$ &
		& $0.43$ & \\

		\cmidrule(lr){1-10}

		\textsf{Berk. C.}
		& $6.50$ & {\color{DarkGray}{$(+87\%)$}}
		& $1.19$ & {\color{DarkGray}{$(+372\%)$}}
		&
		& $6.39$ & {\color{DarkGray}{$(+76\%)$}}
		& $1.08$ & {\color{DarkGray}{$(+188\%)$}} \\

		&							   & {\color{DarkGray}{$(+124\%)$}}
		&							   & {\color{DarkGray}{$(+322\%)$}}
		&
		&							   & {\color{DarkGray}{$(+109\%)$}}
		&							   & {\color{DarkGray}{$(+152\%)$}} \\

		\textsf{Berk. H.3}
		& $9.36$ & {\color{DarkGray}{$(+169\%)$}}
		& $0.84$ & {\color{DarkGray}{$(+234\%)$}}
		&
		& $8.75$ & {\color{DarkGray}{$(+140\%)$}}
		& $0.74$ & {\color{DarkGray}{$(+96\%)$}} \\

		&							   & {\color{DarkGray}{$(+222\%)$}}
		&							   & {\color{DarkGray}{$(+199\%)$}}
		&
		&							   & {\color{DarkGray}{$(+186\%)$}}
		&							   & {\color{DarkGray}{$(+72\%)$}} \\

		\textsf{Berk. H.50}
		& $12.31$ & {\color{DarkGray}{$(+254\%)$}}
		& $0.35$ & {\color{DarkGray}{$(+39\%)$}}
		&
		& $12.01$ & {\color{DarkGray}{$(+230\%)$}}
		& $\mathbf{0.30}$ & {\color{DarkGray}{$(-19\%)$}} \\

		&							   & {\color{DarkGray}{$(+323\%)$}}
		&							   & {\color{DarkGray}{$(+24\%)$}}
		&
		&							   & {\color{DarkGray}{$(+293\%)$}}
		&							   & {\color{DarkGray}{$(-29\%)$}} \\

		\expg
		& $4.15$ & {\color{DarkGray}{$(+19\%)$}}
		& $3.83$ & {\color{DarkGray}{$(+1425\%)$}}
		&
		& $5.80$ & {\color{DarkGray}{$(+59\%)$}}
		& $14.05$ & {\color{DarkGray}{$(+3638\%)$}} \\

		&							   & {\color{DarkGray}{$(+43\%)$}}
		&							   & {\color{DarkGray}{$(+1265\%)$}}
		&
		&							   & {\color{DarkGray}{$(+90\%)$}}
		&							   & {\color{DarkGray}{$(+3179\%)$}} \\

		\ken $\msp$ \sf T.
		& $4.58$ & {\color{DarkGray}{$(+32\%)$}}
		& $\mathbf{0.23}$ & {\color{DarkGray}{$(-8\%)$}}
		&
		& $5.04$ & {\color{DarkGray}{$(+39\%)$}}
		& $0.39$ & {\color{DarkGray}{$(+5\%)$}} \\

		&							   & {\color{DarkGray}{$(+58\%)$}}
		&							   & {\color{DarkGray}{$(-18\%)$}}
		&
		&							   & {\color{DarkGray}{$(+65\%)$}}
		&							   & {\color{DarkGray}{$(-8\%)$}} \\

		\cmidrule(lr){1-10}

		\rand
		& $4.01$ & {\color{DarkGray}{$(+15\%)$}}
		& $6.48$ & {\color{DarkGray}{$(+2478\%)$}}
		&
		& $3.86$ & {\color{DarkGray}{$(+6\%)$}}
		& $6.25$ & {\color{DarkGray}{$(+1561\%)$}} \\

		&							   & {\color{DarkGray}{$(+38\%)$}}
		&							   & {\color{DarkGray}{$(+2207\%)$}}
		&
		&							   & {\color{DarkGray}{$(+26\%)$}}
		&							   & {\color{DarkGray}{$(+1357\%)$}} \\
		\cmidrule(lr){1-10}

		\sf MPH
		& $\mathbf{9.92}$ &
		& $0.15$ &
		&
		& $\mathbf{9.94}$ &
		& $0.24$ & \\

		\ken $\msp$ \sf P.3
		& $14.77$ & {\color{DarkGray}{$(+49\%)$}}
		& $0.32$ & {\color{DarkGray}{$(+106\%)$}}
		&
		& $14.84$ & {\color{DarkGray}{$(+49\%)$}}
		& $0.30$ & {\color{DarkGray}{$(+25\%)$}} \\

		\ken $\msp$ \sf P.50
		& $21.48$ & {\color{DarkGray}{$(+117\%)$}}
		& $\mathbf{0.10}$ & {\color{DarkGray}{$(-36\%)$}}
		&
		& $21.57$ & {\color{DarkGray}{$(+117\%)$}}
		& $\mathbf{0.15}$ & {\color{DarkGray}{$(-40\%)$}} \\

    \bottomrule
    \end{tabular}
    }
\caption{Perplexity benchmark results reporting average number of bytes per gram ($\bpg$) and micro seconds per query ($\mpq$) using modified Kneser-Ney $5$-gram language models built from $\euro$ and $\yahoo$ counts.
\textsf{Berk.} is a short-hand name for {\berk}.}
\label{tab:perplexity}
\vspace{-0.5cm}
\end{table}

\parag{Perplexity benchmark}
Besides the efficient indexing of frequency counts, our data structures can also be used to map $n$-grams to language model probabilities and backoffs. As done by $\ken$, we also use the \emph{binning} method~\cite{BF06} to quantize probabilities and backoffs and allowing any quantization bits ranging from $2$ to $32$. Uni-grams values are stored unquantized to favour query speed: as vocabulary size is typically very small compared to the number of total $n$-grams, this has a minimal impact on the space of the data structure.
Our trie implementation is \emph{reversed} as to permit a more efficient computation of sentence-level probabilities, with a \emph{stateful} scoring function that carries its state on from a query to the next, as similarly done by $\ken$ and $\berk$.

For the perplexity benchmark we used the standard query dataset publicly available at \url{http://www.statmt.org/lm-benchmark},
that contains \num{306 688} sentences, for a total of \num{7 790 011} tokens~\cite{Chelba14}.
We used the utilities of $\expg$ to build modified Kneser-Ney~\cite{CG96,CG99} $5$-gram language models from the counts of $\euro$ and $\yahoo$ that have an OOV (out of vocabulary) rate of, respectively, $16\%$ and $1.82\%$ on the test query file.
As $\expg$ only builds quantized models using $8$ quantization bits for both probabilities and backoffs, we also use this number of quantization bits for our tries and $\ken$ trie.
For all data structures, $\berk$ truncates the mantissa of floating-point values to $24$ bits and then stores indices to distinct probabilities and backoffs.
$\rand$ was build, as already said, with the default parameters recommended in the documentation.

\return
Table~\ref{tab:perplexity} shows the results of the benchmark.
As we can see, the $\timeopt$ data structure is as fast as the $\ken$ trie while being more than $30\%$ more compact on average, whereas the $\spaceopt$ variant \emph{doubles} the space gains with negligible loss in query processing speed ($13\%$ slower).
We instead significantly outperform all other competitors in both space and time, including the $\berk$ \textsf{H.3} variant.
In particular, notice that our index is also smaller than the one of $\rand$ which is randomized and, therefore, less accurate.
The query time of $\berk$ \textsf{H.50} is smaller on $\yahoo$; however, it also uses from $3$ up to $4$ times the space of our tries.

\return
The last two rows of the table are dedicated to the comparison of our $\MPH$ table with $\ken$ \textsf{PROBING}.
While our data structure stores quantized probabilities and backoffs, $\ken$ stores uncompressed values for all orders of $N$. We found out that storing unquantized values results in indistinguishable differences in perplexity while unnecessarily increasing the space of the data structure, as it is apparent in the results.
The expensive hash key recombinations necessary for random access are avoided during perplexity computation for the left-to-right nature of the query access pattern.
This makes, not surprisingly, a linear probing implementation actually faster, by $38\%$ on average, than a minimal perfect hash approach when a large multiplicative factor is used for tables allocation (\textsf{P.50}). The price to pay is, however, the double of the space.
On the other hand, the \textsf{P.3} variant is larger (by $50\%$) and slower (by $60\%$ on average).

\section{Fast Estimation}\label{sec:estimation}
The problem we tackle in this section of the paper is the one of \emph{estimating} a modified \emph{Kneser-Ney} language model (see the background Section~\ref{subsec:kneser-ney}), i.e., computing the probability and backoff penalty for every $n$-gram, $1 \leq n \leq N$, extracted from a large textual source.

\subsection{Improved construction: the 1-Sort algorithm}\label{subsec:estimation:1-sort}

\begin{wrapfigure}{l}{0.4\textwidth}
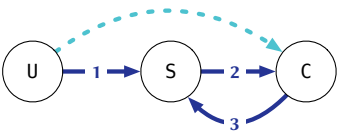

\centering
\includegraphics[scale=0.4]{{{figures/optimization}}}
\caption{Sorting passes performed between $N$-grams: unsorted (\textsf{U}), suffix-sorted (\textsf{S}) and context-sorted (\textsf{C}). Solid arrows describe the path followed by the $\ts$ algorithm; the dashed arrow the one followed by the $\os$ algorithm.
\label{fig:opt}}
\end{wrapfigure}

%Here we describe our main result for the aforementioned problem that is an estimation algorithm for unpruned, modified, Kneser-Ney language models which substantially improve upon the I/O efficiency of $\ts$ by requiring \emph{only one} sorting in external memory.

From the description given in Section~\ref{subsec:estimation:pre} we observe that the running time of $\ts$ is dominated by the cost of sorting in external memory, which is paid \emph{three times} in total: (1) from extraction order (unsorted) to suffix order, (2) from suffix order to context order and then (3) from context order to (again) suffix order. This round-trip is the performance bottleneck of $\ts$ and it is graphically represented in Figure~\ref{fig:opt}.
The natural question is whether it is possible to avoid the round-trip and perform the whole estimation by exploiting a single ordering over the $N$-gram strings.
This section of the paper answers positively to such question by designing an algorithm that requires only one sorting step in external memory.

\return
As an overview, the $\os$ algorithm we are going to describe performs three steps:
(1) counting (Section~\ref{subsec:estimation:1-sort:counting});
(2) adjusting counts (Section~\ref{subsec:estimation:1-sort:adjusting});
(3) in a single, last, pass: normalization and interpolation (Section~\ref{subsec:estimation:1-sort:normalization-interpolation}), joining and index construction (Section~\ref{subsec:estimation:1-sort:joining-indexing}).

%\begin{enumerate}[leftmargin=*]
%\item Counting $N$-grams (Section~\ref{subsec:estimation:1-sort:counting}).
%\item Computing discount coefficients (Section~\ref{subsec:estimation:1-sort:adjusting}).
%\item In a single, last, pass: normalization and interpolation (Section~\ref{subsec:estimation:1-sort:normalization-interpolation}), joining and index construction (Section~\ref{subsec:estimation:1-sort:joining-indexing}).
%\end{enumerate}

In what follows we detail these steps and, thus, show how to save two steps of sorting in external memory.

\subsubsection{Counting}\label{subsec:estimation:1-sort:counting}
This first step is performed similarly to the counting step of $\ts$.
A window of $N$ words slides by one word at a time to scan the input text completely.
We maintain an in-memory block of bytes to accommodate as many $N$-grams as possible, i.e., without taking more space than the amount of RAM specified by the user. Specifically, the block stores records of the form $\langle w_1^N, c(w_1^N) \rangle$, each taking $4N$ bytes for its vocabulary identifiers, plus an 8-byte frequency count.
In order to tell whether an $N$-gram was already seen or not during the scanning of the input, we associate a 4-byte identifier to each distinct $N$-gram by resorting to an open-addressing hash set.
If a cell of the set is not empty and contains the identifier $k \geq 0$, our probe consists in comparing the extracted $N$-gram string with the $4N$ bytes stored in the block starting from the byte at position $k \times (4N + 8)$. If the comparison yields equality, then we increment the corresponding count, otherwise we advance to the next probe position.
If any probed cell is found to be empty, then we write there the next available identifier (equal to the number of distinct seen $N$-grams) and append a new record to the in-memory block.
As soon as we completely fill the block, we use a parallel thread to sort and write it to disk, thus hash deduplication of the text and I/O operations happen simultaneously.

The key difference of this step with respect to the one of $\ts$ lies in the fact that we sort the blocks in \emph{context} order instead of suffix order.
The reason for this choice will become clear as we proceed in the description of the subsequent steps.

\subsubsection{Adjusting counts}\label{subsec:estimation:1-sort:adjusting}
All blocks written to disk by the Counting step are merged together during this step to obtain a single block $B_N$, listing all distinct $N$-grams sorted in context order.
During the process of merging the blocks, we collect the smoothing statistics $t_{n,k}$ in order to use the closed-form estimate of discount coefficients $D_n(k)$, for $k=1,\ldots,4$ (Equation \ref{eq:discount}).

Because smoothing statistics and, thus, discount coefficients, depend on the modified counts of the $n$-grams, the key ingredient we develop in this subsection is a linear-time algorithm that computes the modified counts of all $n$-grams for $1 \leq n < N$ \emph{by scanning the context-sorted block} $B_N$.

Specifically, the records written by the Counting step are merged and accumulated in an in-memory block \emph{block}$[1,m]$ of $m$ records.
When the block fills up, we run the algorithm over the block and then write it to disk. We repeat the process until the whole input $B_N$ is processed completely.
At the end of the process, we use Equation~\ref{eq:discount} to compute the discount coefficients $D_n(k)$.

Before illustrating the algorithm for computing the modified counts over the context-sorted block $B_N$, we first discuss its immediate advantage and then introduce the property of $N$-grams that the algorithm exploits.
Recall that $\ts$ computes the modified counts of the $n$-grams by scanning $B_N$ as sorted in \emph{suffix order}.
Because the next step of estimation is normalization and it requires context order, computing discount coefficients \emph{directly} over the strings sorted in context order has the benefit of \emph{avoiding to sort from suffix to context}.
We are, therefore, eliminating the sorting step 2 shown in Figure~\ref{fig:opt}.

\parag{Exploiting the completeness of the $N$-gram strings}
First of all, observe that since estimation is done without pruning by assumption \emph{and} $N$-grams are extracted using a window of size $N$ that slides by one word at a time, \emph{the strings in $B_N$ cover the input text completely}.
This means that \emph{all} the substrings of length $1 \leq n < N$ of each $N$-gram occur as substrings of some other $N$-gram in $B_N$.
Refer to Figure~\ref{fig:context-sorted-block} and consider the first $5$-gram \textsf{ABAAC} in the context-sorted block.
For example, we know that its sub-string \textsf{BAA} must appear at positions 1, 2 and 3 of some other 5-grams (the ones in position 7, 1 and 2, respectively).
In particular we know that its \emph{prefix} of length 4, i.e., \textsf{ABAA} will be matched at position 2 in some other 5-gram (in this case, the second one, i.e., \textsf{XABAA}).
%We will return to this point later on, in Section~\ref{subsec:estimation:1-sort:joining-indexing}.

%\begin{wrapfigure}{l}{0.5\textwidth}
\begin{figure}
\centering
\includegraphics[scale=0.6]{{{figures/context-sorted-block}}}
\caption{The left extensions (words in blue) of \textsf{AC} must be found in the region highlighted by the light green rectangle, that is the run of entries whose context of length 1 is equal to \textsf{A}.
\label{fig:context-sorted-block}}
\end{figure}
%\end{wrapfigure}

\return
This observation means that all lower-order $n$-grams are \emph{implicitly} contained in the single source block $B_N$.
Two important facts are direct consequences of this property.
\begin{enumerate}[leftmargin=*]
\item  A sorted scan of the $n$-grams can be performed by just scanning $B_N$, \emph{without} the need of replicating on disk all other $n$-grams, for $1 \leq n < N$.
\item Let $C_{n-1}$ be the context of length $n-1$ of an $n$-gram $w_1^n$. The number of distinct left extensions, i.e., the distinct words appearing to the left of $w_1^n$, can be computed by \emph{scanning the $N$-grams whose context of length $n-1$ is equal to} $C_{n-1}$\footnote{Observe that we could compute the left extensions for an $n$-gram by directly scanning the $N$-grams having $w_1^n$ as a context of length $n$. Again, consider the example in Figure~\ref{fig:context-sorted-block}. We could scan the $N$-grams in position 7 and 8 to compute the distinct left extensions (words in blue) of the bigram \textsf{AC}, instead of the ones in position 1, 2, 3 and 4.
The problem with this approach is that we would not be able to compute the wanted quantity for $(N-1)$-grams because, obviously, a context of length $N-1$ can not be extended to the left.
Moreover, consider the first $5$-gram \textsf{ABAAC}.
Since interpolation produces the probabilities for all its suffixes, i.e., for \textsf{C}, \textsf{AC}, \textsf{AAC} and \textsf{BAAC}, we need the modified counts for \emph{these} suffixes and \emph{not} for its contexts \textsf{A}, \textsf{AA}, \textsf{BAA} and \textsf{ABAA} that we could have computed with the other approach.}.
\end{enumerate}
By exploiting these two properties, we now explain the linear-time algorithm for computing the distinct left extensions in context order.

\parag{Computing distinct left extensions in context order}
For ease of explanation, let us consider an $N$-gram $w_1^N$ as composed by three pieces, in order: $P$, $C_{n-1}$ and $w_N$, where $C_{n-1}$ is the context of length $n-1$ and $P$ is the remaining prefix.
Our aim is to compute the number of distinct words $w_{N-n-1}$ \emph{to the left of} the $n$-gram $C_{n-1}w_N$, because this quantity will be its adjusted count, i.e., $a(C_{n-1}w_N)$.
Since $B_N$ is sorted in context order, the entries $P C_{n-1}$ are consecutive for every context $C_{n-1}$, but entries $C_{n-1} w_N$ could not (these entries $C_{n-1}w_N$ are clearly consecutive in suffix order).
However, from fact (2), we know that \emph{every left extension must necessarily appear to the left of the context $C_{n-1}$}, and thus we need to only scan the entries having context $C_{n-1}$.

The quantity $a(C_{n-1}w_N)$ is computed using a direct-address table of size $\Theta(V)$, called \emph{statistics} in the pseudo code shown in Figure~\ref{alg:update}, in which we store, for each distinct $w_N$, the last seen left word (\emph{left}) and the number of distinct left words seen so far (\emph{count}).

As long as context $C_{n-1}$ remains the same during the scan of the block, we look at the table entry corresponding to $w_N$ ($right$) and consider its last seen left word: if different from $w_{N-n-1}$ then we increment its count by one and update the last seen left word with the current one; otherwise we do nothing.
This update step takes $O(1)$ worst-case and it is coded in the \textbf{\texttt{update}} function shown in Figure~\ref{alg:update}.
We are sure to count correctly the number of left extensions because left words are seen in sorted order.

\begin{figure}
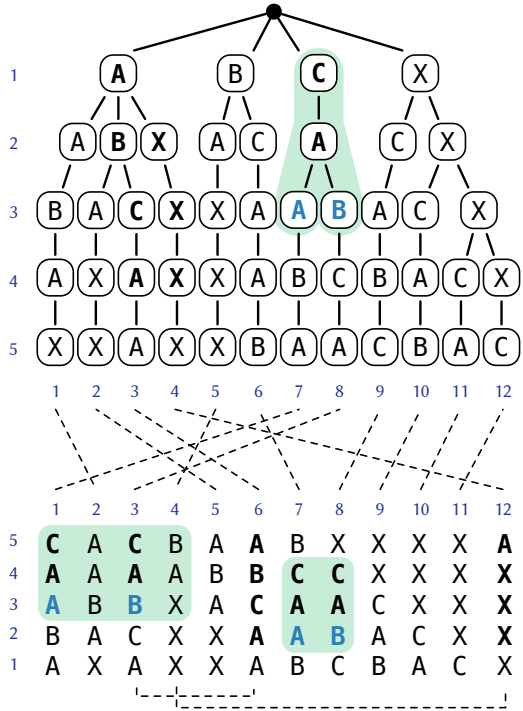

\hspace{-0.5cm}
\includegraphics[scale = 0.6]{{{figures/relation}}}
\caption{The $5$-gram block sorted in context order of Figure~\ref{fig:context-sorted-block} in relation with its reverse trie representation.
The bottom level of the trie, i.e., $[$X, X, A, X, X, B, A, A, C, B, A, C$]$ is obtained by permuting the \emph{first} words of the strings in the context-sorted block, i.e., $[$A, X, A, X, X, A, B, C, B, A, C, X$]$, according to the lexicographic position of their \emph{last} words, i.e., $[$C, A, C, B, A, A, B, X, X, X, X, A$]$.
The left extensions (words in blue) of \textsf{AC} correspond to the children of \textsf{CA} in the reverse trie representation.
}
\label{fig:relation}
\end{figure}

Figure~\ref{fig:context-sorted-block} shows an example for the bigram \textsf{AC}. In this case we have $C_{n-1} = \textsf{A}$, thus we need to scan all the (consecutive) $N$-grams having an \textsf{A} as a context of length 1. These $N$-grams are the ones spanned by the light green rectangle in Figure~\ref{fig:context-sorted-block}.
In this example, \textsf{AC} can be extended to the left with words \textsf{A} and \textsf{B}, as depicted in blue in the picture, thus $a(\textsf{AC}) = 2$.
Also observe that these two words, \textsf{A} and \textsf{B}, correspond to the children of the bigram \textsf{CA} in the \emph{reverse} trie representation of the block shown in the upper part of the Figure~\ref{fig:relation}.
%The trie stores the strings in suffix order.
%In other terms, the node spelling out the bigram \textsf{CA} will store two pointers: one for \textsf{A} and one for \textsf{B}.
We will return to this point when we will discuss how to lay out efficiently the reverse trie, in Section~\ref{subsec:estimation:1-sort:joining-indexing}.

\begin{figure}[t]

    \subfloat[]{
        \scalebox{0.9}{
            \begin{minipage}{.5\textwidth}
            \input{compute-left-extensions}
            \end{minipage}
        }
    }
    \subfloat[]{
        \scalebox{0.9}{
            \begin{minipage}{.5\textwidth}
            \input{update}
            \end{minipage}
         }
    }

  \caption{The \textbf{compute\_left\_extensions} and \textbf{update} functions.
  \label{alg:compute-left-extensions-update}}
\end{figure}

\return

\begin{wrapfigure}{l}{.35\textwidth}
    \centering
    \scalebox{0.9}{\input{was-not-seen}}    
    \caption{The \textbf{\texttt{not\_seen}} function, which checks whether the \textit{right} word was not seen in the current \textit{range}.
    \label{alg:was-not-seen}}
    \vspace{-0.2cm}
\end{wrapfigure}

At the end of the scan of all entries with the same context $C_{n-1}$, it is therefore guaranteed that the table contains the modified counts for all the $n$-grams $C_{n-1}x$.

\return
When the context $C_{n-1}$ changes (Line 7 in the pseudo code of Figure~\ref{alg:compute-left-extensions}), then we would need a fast way to set all counts in the table to zero. Instead, we do not re-initialize the table explicitly that would cost $\Theta(V)$ time, but we associate each context an increasing identifier, as follows.

We store in the table \emph{statistics} an identifier for each distinct word $w_N$, called \emph{range} in the function \textbf{\texttt{not\_seen}} of Figure~\ref{alg:was-not-seen}, that represents the identifier of the range in which the word $w_N$ was last seen.
We also keep track of the current range identifiers in an array $ranges[1,N-2]$.
Now, during the update step we first check the context identifier for the current word $w_N$: if different from the current one, we set its count in the table to zero and update its range identifier accordingly (Figure~\ref{alg:was-not-seen} and Lines 6-8 in the pseudo code of Figure~\ref{alg:update}).

\return
Before concluding, there are two corner cases that we must mention for completeness: the one of $N$-grams and the one of $1$-grams. The former because $N$-grams do not have modified counts, rather their counts are equal to the raw frequency counts written in the input block $B_N$ (Lines 15-17 in Figure~\ref{alg:compute-left-extensions}).
The latter because their context is empty and we do not have to re-initialize their counts in the table when we switch range (\texttt{if} at Line 5 in the pseudo code in Figure~\ref{alg:update}).

%This concludes the description of the linear-time algorithm that uses $\Theta(V)$ space to compute the modified counts of all $n$-grams over a context-sorted block of $N$-grams.

\parag{Collecting smoothing statistics}
We finally describe how we collect the smoothing statistics $t_{n,k}$ for $k=1,\ldots,4$ by using the introduced algorithm.
For each order $n$, we maintain an array $R[1,4]$, where $R[k]$ will store the quantity $|\{w_1^n : a(w_1^n) = k\}|$.
A trivial solution scans the table used by the algorithm whenever we change context and update the counters accordingly. This approach is clearly infeasible in terms of running time.
Instead, we can update each $R[k]$ in $O(1)$ on-the-fly, during the \textbf{\texttt{update}} function of the algorithm, as follows.
Whenever we increment the occurrence of $w_N$ from $k$ to $k+1$ (Line 11 in Figure~\ref{alg:update}), we just have to check the value of $k$: if $k = 1$ then we only increment $R[1]$; otherwise, if $1 < k \leq 5$ then we increment $R[k]$ and decrement $R[k-1]$ (Lines 12-17 in the pseudo code of Figure~\ref{alg:update}).

Whenever we change context, the local counts accumulated in $R$ are first combined with the global ones in another array $T$ and, then, re-initialized (Lines 9-11 in the pseudo code in Figure~\ref{alg:compute-left-extensions}). Also this re-combining step takes constant time.

Finally, from the computed smoothing statistics we can calculate the discount coefficients $D_n$ using Equation~\ref{eq:discount}.
These are kept in an array $D[1,k]$, one for each order $1 \leq n \leq N$ and $k = 1, 2, 3$.

\subsubsection{Normalization and interpolation}\label{subsec:estimation:1-sort:normalization-interpolation}
The linear-time algorithm that computes the modified counts directly over a context-sorted block of $N$-grams can also be used to calculate pseudo probabilities and backoff values using Equation \ref{eq:norm} and \ref{eq:backoff} respectively, by just scanning $B_N$ and using a direct-address table of size $\Theta(V)$ to read the modified counts.

\begin{figure}[t]

    \subfloat[]{
        \scalebox{0.9}{
            \begin{minipage}{.5\textwidth}
            \input{last}
            \end{minipage}
        }
    }
    \subfloat[]{
        \scalebox{0.9}{
            \begin{minipage}{.55\textwidth}
            \input{write}
            \end{minipage}
         }
    }

  \caption{The \textbf{\texttt{last}} step of estimation and the \textbf{\texttt{write}} function that performs normalization, interpolation and indexing.
  \label{alg:last-write}}
\end{figure}

Refer to the pseudo code in Figure~\ref{alg:last}.
In order to interpolate all different orders, we produce pseudo probabilities and backoffs for all $n$-grams sharing the same context, starting from order 2 up to $N$.
This guarantees that as soon as we compute $u(w_N | w_{N-n-1}^{N-1})$ for $2 \leq n < N$, we can directly interpolate it with $\pr(w_N | w_{N-n}^{N-1})$ that has been already computed.
Therefore, the function \textbf{\texttt{write}} in Figure~\ref{alg:write} normalizes and interpolates all $n$-grams sharing the same context (there are $size$ of them at each iteration of the loop).
We now discuss some details about the pseudo code.

\begin{wrapfigure}{l}{.38\textwidth}
    \centering  
    \scalebox{0.9}{\input{uni-grams.tex}}   
    \caption{Final interpolated probability for the unigram $w_n$. The denominator for the quantity $u$ is equal to the number of bigrams in the text, called $m_2$.
    \label{alg:unigrams}}
\end{wrapfigure}

We accumulate the interpolated probabilities of the $n$-grams sharing the same context in an array called \emph{probabilities} and read them sequentially when needed to perform interpolation by using another array of \emph{offsets}.

The body of the function consists in three loops.
The loop in the Lines 4-6 calculates the numerator of the backoff for the context. The one in the Lines 8-11 calculates the denominator for normalized probabilities and backoffs. Finally, the one in the Lines 13-20 calculates the interpolated probabilities.
As already observed, the case for the $N$-grams in Line 22 is identical to the general case for $n < N$ with the only difference that the $N$-grams' counts are not modified but are the raw occurrence counts as seen in the text (see Figure~\ref{alg:indexing-N-grams}).
Finally, for ease of presentation, the Line 17 assumes that the unigrams' probabilities are stored in the array \emph{probabilities}$[1]$. Actually, a unigram probability $\pr(w_n)$ can be computed in $O(1)$ when needed as illustrated in the pseudo code in Figure~\ref{alg:unigrams}, therefore we do not need to buffer them into memory.

\return
In conclusion, normalization and interpolation are carried on as explained for the $\ts$ algorithm (see Section~\ref{3-sort}), but \emph{without} requiring two separate sorting passes over the $N$-gram strings.
Another crucial difference is that the two phases are performed during the same scan of only one block, i.e., $B_N$, and we do not need to jointly iterate through $N$ distinct files, one for each value of $n$, as done by $\ts$.
The net result is that we \emph{avoid to sort from context to suffix} in order to perform interpolation, thus eliminating the sorting step 3 of Figure~\ref{fig:opt}.
Summing up, given that we have formerly shown how to save the sorting from suffix to context too (Section~\ref{subsec:estimation:1-sort:adjusting}), we have completely eliminated the round-trip of $\ts$ mentioned at the beginning of Section~\ref{subsec:estimation:1-sort}.

%\begin{figure}[t]
%  \begin{minipage}{.41\textwidth}
%  \scalebox{0.95}{  
%   \input{last}
%  }
%  \end{minipage}
%  \begin{minipage}{.58\textwidth}
%  \scalebox{0.85}{
%   \input{write}
%  }
%  \end{minipage}
%  \hspace{-2cm}
%  
%  \caption{The main loop of the \textbf{\texttt{last}} step of estimation and the \textbf{\texttt{write}} function that performs normalization and interpolation.
%  \label{alg:last-write}}
%\end{figure}

\subsubsection{Joining and indexing}\label{subsec:estimation:1-sort:joining-indexing}
We now show how to perform the two remaining steps of estimation, i.e., first, the joining of probabilities with backoff values and, second, the building of the reverse trie data structure during the same pass.

We recall that the output of this last step is the compressed, static, trie index that maps the extracted $n$-gram strings to their Kneser-Ney probabilities and backoffs, described in Section~\ref{subsec:indexing:core}.
In particular, it is the \emph{reverse} trie variant, such as the one depicted in Figure~\ref{fig:relation}, because it optimizes the left-to-right pattern of lookups performed by perplexity scoring (see the perplexity benchmark in Section~\ref{subsec:indexing:overall}).

\return
For this problem, we exploit the property already mentioned in Section~\ref{subsec:estimation:1-sort:adjusting}, that is: \emph{every $N$-gram prefix of length $N-1$ must be matched at position $2$ in some other $N$-gram}.
This property gives us two important guarantees.
\begin{enumerate}[leftmargin=*]
\item The first $N-1$ levels of the reverse trie can be built by streaming through the $N$-grams in context order.
\item Backoffs are emitted in suffix order.
\end{enumerate}

In the following we exploit the first guarantee to build the reverse trie data structure and the second one to perform joining of probabilities with backoffs.
By looking at Figure~\ref{fig:relation} that shows the context-sorted block of Figure~\ref{fig:context-sorted-block} in relation to its reverse trie representation, we can graphically visualize these two guarantees.
Let us discuss them separately.

\return

Regarding guarantee (1), we can immediately see that the first 4 levels of the trie are indeed the contexts of length 4 of the $5$-grams in the context-sorted block.
For example, the prefix of length 4 of \textsf{ACBAC}, i.e., \textsf{ACBA}, is found in the $6$-th string; the one of \textsf{XXXAB} in the $12$-th string instead (following the dashed lines at the bottom of Figure~\ref{fig:relation}).
Notice that we \emph{always} find the match at position 2, thus the first 4 levels of the trie store such prefixes.
In general we have that: the first $N-1$ levels of the reversed trie \emph{are the prefixes} of size $N-1$ of the context-sorted $N$-grams and can be, therefore, efficiently built directly from the context-sorted $N$-grams \emph{without} having to sort the $N$-grams in suffix order.

Regarding guarantee (2), consider the first 5-gram \textsf{ABAAC}.
Since interpolation produces the probabilities for all the suffixes, i.e., for \textsf{C}, \textsf{AC}, \textsf{AAC} and \textsf{BAAC}, \emph{we compute the backoffs for their contexts}, i.e., $b(\varepsilon)$, $b(\textsf{A})$, $b(\textsf{AA})$ and $b(\textsf{BAA})$ respectively, which appear in sorted order in the block.
Refer to Figure~\ref{fig:KN-prob} too for a graphical example.
Backoffs are, therefore, computed in suffix order and can be written directly in the corresponding trie nodes.

\return
Now that we know how to efficiently build the first $N-1$ levels of the reversed trie and perform joining, we are left to consider two problems:
first, how to handle the bottom level of the trie and, second, how to write the interpolated probabilities in the nodes of the trie.
In fact, notice that:
regarding the first problem, we can not build the bottom level of the trie directly because a context of length $N-1$ does not extend to the left;
regarding the second problem, interpolation produces the probabilities for the \emph{suffixes} but we rather would need the ones for the \emph{contexts} in order to write them in the trie as we can do for the backoffs.
We clarify this latter point by continuing the example for \textsf{ABAAC}.
We interpolate its constituent $n$-grams in the following (suffix) order: \textsf{C}, \textsf{AC}, \textsf{AAC}, \textsf{BAAC} and \textsf{ABAAC}, but we would actually need the probabilities for the contexts \textsf{A}, \textsf{AA}, \textsf{BAA} and \textsf{ABAA}, in order to write them in the suffix trie (as done for the backoffs).

\begin{wrapfigure}{l}{.45\textwidth}
    \centering
    \subfloat[]{
        \scalebox{0.9}{\input{indexing-ngrams}}
    }
    
    \subfloat[]{
        \scalebox{0.9}{\input{indexing-N-grams}}
    }
    
  \caption{The pseudo code that illustrates how to perform indexing, for the case $n < N$ in (a) and for the case $n = N$ in (b). The two listings complete the pseudo code in Figure~\ref{alg:write}.
  \label{alg:ab}}
  \vspace{-0.3cm}
\end{wrapfigure}

\parag{Exploiting the relation between context and suffix order}
To efficiently solve these two remaining problems, we exploit the following property that establishes the relation between context and suffix order:
\emph{A context-sorted block can be sorted efficiently in suffix order by considering the order on the last word only}, because the prefixes of length $N-1$ are already sorted.

In turn, this property implies that:
\emph{The bottom level of the trie can be built by placing the first words of the strings of the context-sorted block in the lexicographic positions of their last words}.
Thanks to this property, although the algorithm operates over the strings sorted in context order, it is still able to efficiently lay out the strings in suffix order.

\return
The relation is depicted in Figure~\ref{fig:relation} by the dashed lines linking the context-sorted $5$-grams with the corresponding root-to-leaf paths in the reverse trie.
For example, consider the first $5$-gram \textsf{ABAAC}. We know that such string will terminate with \textsf{A} (first word) in the bottom level of the trie. The position at which we have to place this first word in the bottom level is the lexicographic position of the last word, i.e., the \textsf{C}. Since the lexicographic position of the \textsf{C} is 7 within all the last words of the $5$-grams (4 \textsf{A}s and 2 \textsf{B}s first), \textsf{A} is placed in position 7 in the last level of the trie (follow the dashed line from position 1 in the context-sorted block to position 7 in the trie).

\newpage

\return
In order to place word identifiers and probabilities in correct position, we use a \emph{count-indexing technique}.
%, instead of storing the permutation explicitly that would cost $m_N \lceil \log m_N \rceil $ bits of space, where $m_N$ is the total number of $N$-grams.
For each vocabulary word, we maintain the number of times it appears \emph{as last word} of an $N$-gram in a direct-address table of size $\Theta(V)$.
Prefix-summing such counts (shifted by one position to the right) gives us in $O(1)$, for each distinct word identifier $w_N$, the position in the array, that represents the bottom level of the trie, at which we have to write the first occurrence of $w_N$.
Given such position, we write the integer $w_N$ in $O(1)$ and increment the position in the table by one.
Notice that this is the same procedure used by counting sort, thus the correctness of the approach follows automatically (see Section 8.2 of~\cite{CLRS}).
It only requires $V$ integer counters, that we store in an array \emph{positions}$[1,N]$.

\return
Let us consider a complete example. Refer to Figure~\ref{fig:relation} and the pseudo code in Figure~\ref{alg:indexing-N-grams}.
For the unigrams \textsf{A}, \textsf{B}, \textsf{C} and \textsf{X}, we count how many times they appear as last words of the $N$-grams and we obtain the following counts $[4, 2, 2, 4]$, because \textsf{A} and \textsf{X} appear 4 times each, while \textsf{B} and \textsf{C} appear twice each.
Now we prefix sums such counts\footnote{And also sum 1 because our examples use 1-based indexes.}, obtaining $[5, 7, 9, 13]$, and we shift them one position to the right, obtaining the following initial \emph{positions}$[5][4]$ = $[1, 5, 7, 9]$.

Now, consider the first $5$-gram in the context-sorted block, i.e., \textsf{ABAAC}. Since its last word is \textsf{C}, we look at its initial position in the array, which is 7, and we know that we have to place its first word, \textsf{A}, at position 7 in the last level of the trie.
This is done in Line 9 of the pseudo code.
%, where the trie is assumed to be represented as an array of \emph{levels}, i.e., \emph{levels}$[N]$ is the last level of the trie.
As a matter of fact, the 7-th string in the reverse trie of Figure~\ref{fig:relation} is exactly \textsf{ABAAC}.
Then, we know that the second occurrence of \textsf{C} (last word of \textsf{ACBAC}) will give us position $7+1=8$. Thus, we will write an \textsf{A} in position 8.
Let us now consider the second $N$-gram, i.e., \textsf{XABAA}.
The position associated to \textsf{A} is 1, so we have to write the first word \textsf{X} at position 1.
We repeat the process for all the $N$-grams in the context-sorted block:
following the dashed lines of Figure~\ref{fig:relation}, it is easy to see that the last level of the trie can be built correctly by the introduced algorithm.
The corresponding pseudo code is illustrated in Figure~\ref{alg:indexing-N-grams} and it represents the case for $n = N$ in the \textbf{\texttt{write}} pseudo code in Figure~\ref{alg:write} (Line 22).

\return
The same technique is also used to place the final probabilities in the correct trie nodes for all orders $1 < n \leq N$.
Let us consider a full example for $n = 2$ in order to explain how this is possible.
For the unigrams \textsf{A}, \textsf{B}, \textsf{C} and \textsf{X}, we obtain the following counts $[3, 2, 1, 2]$. In fact, although \textsf{A} appears 4 times, it only appears in 3 distinct contexts, i.e., to the right of the bigrams \textsf{AA}, \textsf{BA} (that appears twice) and \textsf{XA}. Instead, \textsf{B} appears twice: once to the right of \textsf{AB} and to the right of \textsf{CB}.
As done before, prefix-summing and shifting the counts, we obtain the initial \emph{positions}$[2][4]$ = $[1, 4, 6, 7]$.
Now, consider the first $5$-gram \textsf{ABAAC}. When we produce the final interpolated probability for \textsf{AC}, we have to write it in the second level of the trie in position 6 as given by corresponding counter in the array. Again, we can immediately verify that the (6+1)-th root-to-leaf path in the trie is the one spelling out \textsf{CA}.
For the second $5$-gram \textsf{XABAA}, instead, we have to write the probability of \textsf{AA} at position 1 in the second level of the trie.

The examples above can be easily extended to any other order $2 < n \leq N$.
In this case, the corresponding pseudo code is illustrated in Figure~\ref{alg:indexing-ngrams} and it completes the \textbf{\texttt{write}} function coded in Figure~\ref{alg:write} (Line 20).

\return
Finally, we also have to write the pointers for each node of the trie. However, observe that a pointer represents the number of successors of a given $n$-gram, thus pointers are \emph{the same as the modified counts}.
Therefore, pointers require no extra effort (and are not shown in the pseudo code for simplicity).

\subsection{Experiments}\label{subsec:estimation:exp}
The experiments we will show in this subsection have the purpose of first analyzing the running time of our solution, i.e., the $\os$ algorithm, then of introducing optimizations and, finally, of considering the comparison against the $\ts$ approach.

\parag{Datasets}
We performed our experiments using the following textual collections in the English language.
\begin{itemize}[leftmargin=*]
\item {\BW} is the concatenation of all the news files contained in the \textsf{training} directory of the dataset described in~\cite{Chelba14} and publicly available at: \url{http://www.statmt.org/lm-benchmark},
\item {\WP} is a recent Wikipedia dump, collected from October to December 2017 and publicly available at: \url{https://dumps.wikimedia.org/enwiki/latest}.
\item {\CW} is a sampling of 5 million pages drawn from the ClueWeb 2009 TREC Category B test collection, consisting of English web pages crawled between January and February 2009, available at: \url{http://www.lemurproject.org/clueweb09}.
\end{itemize}

From each dataset we removed all non-ASCII characters and markup tags.
We use the standard value of $N=5$ in every experiment, as already done for experiments presented in Section~\ref{subsec:indexing:exp}.
The datasets are of increasing size, reported as the number of $n$-grams in Table~\ref{tab:estimation:datasets_stats}: this will be useful to show the behaviour of our solution by varying the size of the input.

\begin{table}[t]
  \centering
  \scalebox{0.9}{
    \begin{tabular}{c r r r}
        \toprule
        $n$ & $\BW$ & $\WP$ & $\CW$ \\

% \cmidrule(lr){2-3}
% \cmidrule(lr){4-5}
% \cmidrule(lr){6-7}

\midrule

1 & \num{2438616}
  & \num{5681625}
  % & \num{8769460}   \\
  & \num{4291588}   \\

2 & \num{43179094}
  & \num{141639447}
  % & \num{111081941} \\
  & \num{236626867} \\

3 & \num{203793974}
  & \num{587261939}
  % & \num{547305695} \\
  & \num{977038965} \\

4 & \num{427172514}
  & \num{1115647651}
  % & \num{1324876160} \\
  & \num{1710815581} \\

5 & \num{588390914}
  & \num{1463820688}
  % & \num{2051542820} \\
  & \num{2129634982} \\

\midrule

total            & \num{1264975112}
                 & \num{3314051350}
                 % & \num{4043576076}
                 & \num{5058407983}
                 \\

        \bottomrule
    \end{tabular}
   }
\caption{Number of $n$-grams for the datasets used in the experiments.}
\label{tab:estimation:datasets_stats}
\end{table}

\parag{Experimental setting and methodology}
All experiments have been performed on a machine with 4 Intel \textsf{i7-7700} cores clocked at 3.6 GHz, with 64 GB of RAM \textsf{DDR3}, running Linux 4.4.0, 64 bits. RAM is clocked at 2.133 GHz.
%Hardware caches have the following sizes: 256 KB (\textsf{L1}), 1 MB (\textsf{L2}) and 8 MB (\textsf{L3}).
%L1d cache:             32K
%L1i cache:             32K
%L2 cache:              256K
%L3 cache:              8192K
The machine is equipped with a mechanical disk of 3 TB \textsf{WDC WD30EFRX-68E}, with standard page size of 4 KB.

%To measure the R/W speed of the disks, we create a binary file of 1GB and measure the time for writing and reading the file. We make sure to avoid caching effect of modern filesystems by using the \textsf{sync} command and clearing caches with \path{/proc/sys/vm/drop_caches}.
%With this experiment we obtain a R/W speed of 140 MB/s. Exploiting caching effects makes the experiment run approximately 57 times faster, for a R/W speed of 8 GB/s. Thus it is particularly important to read/write from/to the same file because its pages are likely to be cached. We will return to this point later on as we discuss the experiments.

We implemented the $\os$ algorithm in standard \textsf{C++14}, whose code is available freely at \url{https://github.com/jermp/tongrams}.
As our competitor, we use the \textsf{C++} implementation of $\ts$ as provided by the authors of~\cite{HPCK13} and available at \url{http://kheafield.com/code/kenlm}. We refer to this implementation as $\ken$, which is the lead toolkit for language modeling~\cite{Heafield11}.
%As a matter of fact, $\ken$ provides the fastest estimation algorithm, significantly outperforming the previous approaches~\cite{HPCK13} as reported in Section~\ref{subsec:estimation:pre}.
%This is also confirmed by other recent experiments, showing $\ken$ to be up to $10\times$ faster to build for the typical values of $n \leq 5$ than approaches based on compressed suffix trees~\cite{shareghi2016fast}.

Both implementations were compiled with \textsf{gcc} 5.4.0, using the highest optimization setting, i.e., with compilation flags \texttt{-O3} and \texttt{-march=native}.

\subsubsection{Preliminary analysis}\label{subsec:estimation:pre-exp}
As a first set of experiments we show the running time of our algorithm by varying the amount of internal memory and by inspecting the CPU and I/O activity.

\begin{figure}
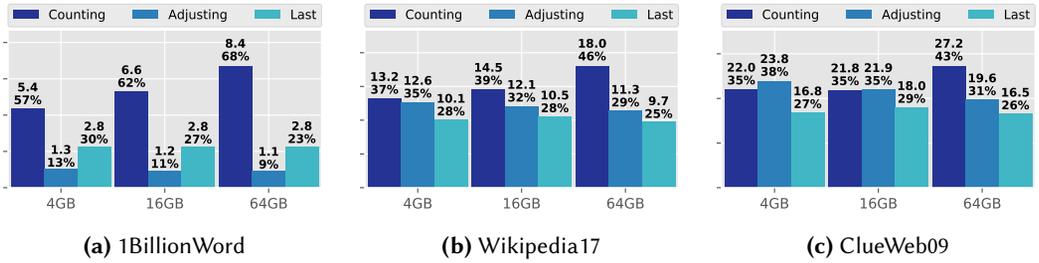

    \subfloat[$\BW$]{
    \includegraphics[scale=0.53]{{{plots/1Billion.varying_RAM}}}
    \label{fig:1Billion.varying_RAM}
    }
    \subfloat[$\WP$]{
    \includegraphics[scale=0.53]{{{plots/wikipedia.varying_RAM}}}
    \label{fig:wikipedia.varying_RAM}
    }
    \subfloat[$\CW$]{
    \includegraphics[scale=0.53]{{{plots/clueweb.varying_RAM}}}
    \label{fig:clueweb.varying_RAM}
    }   
    \caption{Time in minutes spent at each step of estimation by using different amounts of internal memory.}
    \label{fig:varying_RAM}
\end{figure}

\parag{Varying the amount of internal memory}
We show the running time of our algorithm at each step of estimation, by varying the allowed amount of internal memory among 4 GB, 16 GB and the maximum available RAM, 64 GB.
This experiment aims at showing what steps are the most expensive and fix the amount of internal memory that we will use for the subsequent analysis.
The plots in Figure~\ref{fig:varying_RAM} illustrate the results. Above each bar, we report two numbers: the first indicating the number of minutes spent during the step, the second indicating the percentage with respect to the total running time of the algorithm.
This grand total measures the time of the whole estimation process, i.e., the time it takes from the scanning of the input text to the flushing on disk of the compressed index built over the extracted strings.
Some considerations are in order.

First of all, we can observe that, not surprisingly, the size of the language model has a significant impact not only on the total running time but also on which step becomes the most expensive.
In fact, while on the $\BW$ dataset the Counting and the Last steps contribute for more than $80\%$ of the total running time and the Adjusting step has a quite low impact, the trend changes significantly on the larger datasets. In fact, on $\WP$ and $\CW$ the total running time is almost evenly distributed across the three steps. Notice that, in particular, the time for Adjusting rises significantly.
This is due to the number of $N$-gram blocks written to disk during the Counting step and that are merged together during the Adjusting step.
On the smaller dataset $\BW$, we have relatively few blocks to merge, thus Adjusting is performed quickly.
Clearly, using more internal memory helps in lowering the number of blocks to merge and, thus, reducing the time for Adjusting.

We also observe that the step of Counting and the Last one do not vary much when more memory is available.
Concerning the Counting step, more memory is not useful to lower the running time because using larger hash sets also means sorting larger blocks of $N$-grams.
Indeed, observe that the total running time of Counting (slightly) increases by increasing the amount of memory.
However, as we have discussed above, using more memory for sorting implies fewer of blocks to merge, thus internal memory size has an impact only on the Adjusting step.
%The same holds true for $\ken$.
For the open-address hash set implementation that we use in the Counting step, we experimented with linear probing, quadratic probing and double hashing. No significant difference among the three strategies was observed, thus we prefer linear probing for its better locality of accesses.
%Not surprisingly, this is also the choice made by $\ken$.
Concerning the Last step, we need to scan the merged $N$-gram file once. We use a standard buffered-scan approach using blocks of 64 MB by default. Using larger buffers does not impact the running time.

\return
Since similar observations also hold true for $\ken$, we choose the middle value of 16 GB for all datasets as the quantity of memory we use for all the following experiments.

\begin{figure}
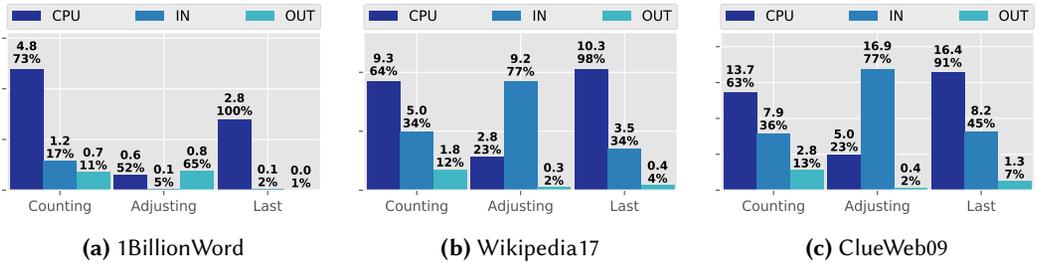

    \subfloat[$\BW$]{
    \includegraphics[scale=0.53]{{{plots/1Billion.CPU_IO_total}}}
    \label{fig:1Billion.CPU_IO_total}
    }
    \subfloat[$\WP$]{
    \includegraphics[scale=0.53]{{{plots/wikipedia.CPU_IO_total}}}
    \label{fig:wikipedia.CPU_IO_total}
    }
    \subfloat[$\CW$]{
    \includegraphics[scale=0.53]{{{plots/clueweb.CPU_IO_total}}}
    \label{fig:clueweb.CPU_IO_total}
    }   
    \caption{Time in minutes spent by CPU computation and I/O activity at each step of estimation.}
    \label{fig:CPU_IO_total}
\end{figure}

\parag{Inspecting CPU and I/O activity}
It is now interesting to quantify the impact that CPU and I/O operations have on the total running time of each step.
Under a different perspective, this analysis is useful to understand \emph{how} disk usage is impacted by the size of the language model.
The plots in Figure~\ref{fig:CPU_IO_total} illustrate such impact, i.e., the time spent by CPU and I/O at each step by using the amount of RAM that we fixed before (16 GB).

Dealing with external memory poses the challenge of trying to avoid CPU idle time by overlapping CPU computation with I/O activity.
For such reason, we use asynchronous threads to handle input/output operations, so that while the CPU is performing internal processing, data is read or written to disk simultaneously~\cite{STXXL}.
This is a feature of particular importance for on-disk programs such as the ones we are considering, given the huge discrepancy in speed of modern processors and (mechanical) disks.
%We point out that a similar technique is also used by our competitor.
Clearly, a perfect overlapping between CPU and I/O time would mean to only pay the maximum of the two.
Consequently, the sum of three percentages for CPU, IN and OUT time for a given step in Figure~\ref{fig:CPU_IO_total}, may exceed $100\%$ because these are handled by different threads.
Let us now consider each step in order.

\return
During the Counting step, while the reader thread is scanning the input and probing the hash set, the writer thread is asynchronously sorting the previous $N$-gram block and flushing it to disk.
While sorting is strictly CPU-bound because it is performed in memory, the scanning of the input text imposes some CPU idle time as apparent for the plots of the larger datasets $\WP$ and $\CW$.
However, probing the hash set and sorting contribute to most of the time spent during the Counting step.
In fact, the plots report that the sum of CPU and IN percentages yields almost the whole running time of Counting, whereas the OUT time is completely overlapped with CPU processing.

\return
The total running time of the Adjusting step is, instead, dominated by the cost of reading the blocks from the disk.
This is no surprise given that multiple input streams are contending the disk for input operations, thus incurring in more disk seeks~\cite{Vitter98}.
As a result, on the larger datasets $\WP$ and $\CW$ we can see the IN time taking $77\%$ of the total: this causes the CPU utilization to drop down to roughly $23\%$, by experiencing idle time.
Indeed, the time taken by the algorithm described in Section~\ref{subsec:estimation:1-sort:adjusting} for computing the left extensions over a context-sorted block,
is very small compared to the overall running time of the step and contributes to a small percentage of the CPU: it is just $0.42$, $1.2$ and $1.8$ minutes on $\BW$, $\WP$ and $\CW$ respectively.
The remaining part of the CPU is spent by iterating through the fetched block of $N$-grams and comparing records during the merging process.

\return
During the Last step, while the reader thread is loading a block from disk, the CPU is processing the previous block. Therefore, we have a good overlap between CPU and reading time from disk. This is possible because disk reads are issued to a single source, i.e., the merged $N$-gram file, thus we avoid the disk seeks experienced during the Adjusting step.
As a result, all time is spent by the CPU.

\newpage

\subsubsection{Optimizing our solution}\label{subsec:estimation:opt-exp}
In this subsection we devise and quantify the impact of one performance optimization for each step of estimation.

\parag{Counting: implementing a parallel radix sort}
In order to lower the total running time of the Counting step, it is important to guarantee a good overlap between input scanning and sorting in order to only pay the maximum of the two latencies and not the sum of the two.
For this reason, we use least-significant-digit (LSD) \emph{radix sort}~\cite{CLRS}, instead of the general-purpose \textsf{std::sort}.
This sorting algorithm is the right choice in our setting because each $N$-gram is a (short) string of exactly $N$ 32-bit numbers, thus $N$ passes of counting sort, i.e., one for each word index $j$, $j = N-1,0,1,\ldots,N-2$, are sufficient (and necessary) to sort a block in context order.
The time complexity to sort a block of $m$ $N$-grams is $\Theta((m+V) \times N)$, which is $\Theta(m \times N)$ given that $V = O(m)$.

Moreover, each step of counting sort on column index $j$ is implemented in parallel, as follows.
Let $K$ be the number of threads used for sorting.
We allocate a table $C[K+1][V]$ of counters, where $C[t+1]$ will store the number of occurrences of each word identifier in the partition of $\Theta(\frac{m}{K})$ records assigned to thread $t$.
Then each thread $t$, for $0 \leq t < K$, runs in parallel and increments by one the entry $C[t+1][i]$ whenever it encounters the word identifier $i$.
Now, prefix-summing the counters by a \emph{column-major scan} of $C$ transforms each entry $C[t][i]$ into the (sorted) position in the output block at which thread $t$ has to write the record having $i$ as its $j$-th word identifier.

Thanks to this strategy and by using all the available cores on our test machine ($K = 4$), the time for the Counting step improves substantially\footnote{During our experimentation, we found out that this parallel implementation of radix sort is also roughly $1.8\times$ faster on average than \textsf{gnu::parallel\_sort}. As an example, to sort an $N$-gram block of 8 GB, the \textsf{gnu::parallel\_sort} takes 30 seconds while our parallel LSD radix sort takes 16.4 seconds.} because sorting $N$-gram blocks becomes completely overlapped with input scanning and probing of the hash set: from 6.6 minutes we pass to 3.5 minutes on $\BW$ ($1.88\times$); from 14.5 to 10 minutes on $\WP$ ($1.45\times$); from 21.8 to 15.8 on $\CW$ ($1.38\times$).

\begin{table}

    \subfloat[$\WP$]{
    \scalebox{0.9}{
    \begin{tabular}{
                              l@{\hspace{10pt}}
                          l@{\hspace{2pt}}
                          r@{\hspace{8pt}}
                          r@{\hspace{2pt}}
                          r@{\hspace{8pt}}
                          r@{\hspace{2pt}}
                          r@{\hspace{8pt}}
                          r@{\hspace{2pt}}
                          r@{\hspace{8pt}}
                          }
    \toprule
    & \multicolumn{2}{@{}l@{}}{\textsf{CPU}}
& \multicolumn{2}{@{}l@{}}{\textsf{IN}}
& \multicolumn{2}{@{}l@{}}{\textsf{total}}
& \multicolumn{2}{@{}l@{}}{\textsf{bytes/gram}}
\\

\cmidrule(lr){2-9}

\textsf{Uncompressed}
& $\mathbf{2.81}$ &
& $9.24$ &
& $12.05$ &
& $28.00$ &
\\

\textsf{FC bit-aligned}
& $5.77$ & \color{DarkGray}{$(0.5\times)$}
& $\mathbf{0.10}$ & \color{DarkGray}{$(97\times)$}
& $5.86$ & \color{DarkGray}{$(2\times)$}
& $\mathbf{9.00}$ & \color{DarkGray}{$(3\times)$}
\\

\textsf{FC byte-aligned}
& $3.94$ & \color{DarkGray}{$(0.7\times)$}
& $1.22$ & \color{DarkGray}{$(8\times)$}
& $\mathbf{5.03}$ & \color{DarkGray}{$(2.4\times)$}
& $11.00$ & \color{DarkGray}{$(2.5\times)$}
\\

    \bottomrule
    \end{tabular}
    }}
    
    \subfloat[$\CW$]{
    \scalebox{0.9}{
    \begin{tabular}{
                              l@{\hspace{10pt}}
                          l@{\hspace{2pt}}
                          r@{\hspace{8pt}}
                          r@{\hspace{2pt}}
                          r@{\hspace{8pt}}
                          r@{\hspace{2pt}}
                          r@{\hspace{8pt}}
                          r@{\hspace{2pt}}
                          r@{\hspace{8pt}}
                          }
    \toprule
    & \multicolumn{2}{@{}l@{}}{\textsf{CPU}}
& \multicolumn{2}{@{}l@{}}{\textsf{IN}}
& \multicolumn{2}{@{}l@{}}{\textsf{total}}
& \multicolumn{2}{@{}l@{}}{\textsf{bytes/gram}}
\\

\cmidrule(lr){2-9}

\textsf{Uncompressed}
& $\mathbf{4.98}$ &
& $16.91$ &
& $21.89$ &
& $28.00$ &
\\

\textsf{FC bit-aligned}
& $9.29$ & \color{DarkGray}{$(0.5\times)$}
& $5.25$ & \color{DarkGray}{$(3\times)$}
& $14.55$ & \color{DarkGray}{$(1.5\times)$}
& $\mathbf{9.75}$ & \color{DarkGray}{$(3\times)$}
\\

\textsf{FC byte-aligned}
& $7.61$ & \color{DarkGray}{$(0.7\times)$}
& $\mathbf{4.23}$ & \color{DarkGray}{$(4\times)$}
& $\mathbf{11.55}$ & \color{DarkGray}{$(2\times)$}
& $11.65$ & \color{DarkGray}{$(2.4\times)$}
\\

    \bottomrule
    \end{tabular}
    }}
    
\caption{The effect of compressing blocks during the Adjusting step, on $\WP$ and $\CW$ datasets. The table reports: the time in minutes spent by computation (\textsf{CPU}), reading from disk (\textsf{IN}) and globally (\textsf{total}) and the average bytes per gram achieved by the different implementations.}
\label{tab:FC}
\end{table}

\parag{Adjusting: compressing $N$-gram blocks}
The high cost of reading the $N$-gram files from disk during the Adjusting step suggests that all efforts spent in enhancing its running time should be devoted in reducing the loading time from disk, because lowering the CPU cost will result in a negligible improvement.
For this reason we compress the $N$-gram blocks created during the Counting step.
Compressing the blocks has the potential of reducing the time spent in reading from disk because more (compressed) $N$-grams are transferred from disk to memory during an input operation.

What we need is a compressed stream representation that supports fast sequential decoding. We adapt a \emph{front-coding} (\textsf{FC})~\cite{Witten99} representation of an $N$-gram block, as follows.
We fix a window size in bytes (64 MB by default, in our implementation) and compress as many records $\langle w_1^N, c(w_1^N) \rangle$ as possible, i.e., as many as can be possibly contained in the window.
When encoding/decoding a window, we maintain the following invariant: a record is either written uncompressed, or compressed with respect to the previous one.
In particular, a record is encoded as a pair $\langle \ell, s \rangle$, where $\ell$ is the number of word identifiers we have to copy from the previous record (in context order) and $s$ is the remaining part of the string (the suffix).
The first record of each window is written uncompressed.

We can use the minimum number of bits or bytes to represent each word identifier and frequency count.
We refer to such strategies as, respectively, \textsf{FC bit-aligned} and \textsf{FC byte-aligned}, whose impact is evaluated in Table~\ref{tab:FC}.
%Since we use $N=5$ in our experiments, we have that $\ell = \lceil \log(N + 1) \rceil = 3$.
%This scheme allows us to support fast iteration over the compressed stream, since we need to copy just from the previously decoded record, thus making an efficient exploitation of the processor cache.
%During the Adjusting step, we read a file one window at a time, with no special boundary-check since windows are all of the same size in bytes, and instantiate an iterator over the compressed stream at the beginning of the window.
As we can see from the data reported in the table, the bit-aligned version offers a $3\times$ space reduction: from 28 bytes per record of the uncompressed version, we pass to an average of $9$ bytes per record on $\WP$ and to $9.75$ bytes per record on $\CW$.
As a net result, the Adjusting step on $\WP$ and $\CW$ runs $2\times$ and $1.5\times$ faster.
Indeed, we can observe that the input time decreases significantly: it is almost $100\times$ smaller on $\WP$ and more than $3\times$ smaller on $\CW$.
However, notice that the CPU time rises as well, roughly $2\times$, due to decoding from a compressed stream: we trade CPU time for less reading from disk.
%This evidence suggests that a faster implementation of front-coding may even improve the total running time.

The byte-aligned version, \textsf{FC byte-aligned}, avoids the many bit-level instructions to decode a record. Not surprisingly, we can see that this strategy is actually faster than the bit-aligned version by $25\%$ on average, while only allowing a slightly worse compression ($2.5\times$ on average compared to $3\times$).
In conclusion, compressing the $N$-gram blocks with byte-aligned front-coding yields an improvement of $2.4\times$ and $1.9\times$ on $\WP$ and $\CW$ datasets, respectively.
Therefore, for the rest of the experiments we use the \textsf{FC byte-aligned} representation of the blocks. 
On the smaller dataset $\BW$, however, compressing the blocks does not yield an appreciable improvement since input time from disk takes a negligible fraction of the total running time of the step (see Figure~\ref{fig:1Billion.CPU_IO_total}).

\parag{Last: processing $N$-gram blocks in parallel}
As discussed in Section~\ref{subsec:estimation:pre-exp}, the Last step of estimation is CPU-bound. Thus, we can use multi-threading to speed up the execution of the step.
If $K$ is the chosen parallelism degree, we use 1 reader thread to load the next $K-1$ blocks from the merged $N$-gram file and $K-1$ worker threads to process these blocks in parallel.
While each worker thread independently executes the step described in Section~\ref{subsec:estimation:1-sort:normalization-interpolation} on its own block (the \textbf{\texttt{last}} function in the pseudo code of Figure~\ref{alg:last}), the reader thread asynchronously loads the next $K-1$ blocks in memory.
The main challenge of this approach lies in computing the partition of each level of the trie that has to be written by a worker thread.
For this problem, we use a 2-step algorithm: in a first phase, each worker thread computes the number of distinct $n$-grams in its own block; in a second phase these counts are combined to obtain the offsets of the global partition of the trie.
Although the first phase is performed in parallel, it has an impact on the achieved scalability.

On our test machine, we have $K=4$, thus we use 3 worker threads and 1 reader thread.
On $\BW$ we reduce the running time from 2.8 to 1.33 minutes ($2.1\times$); on $\WP$ from 10.53 to 6.85 minutes ($1.54\times$); on $\CW$ from 18 to 11.8 minutes ($1.52\times$).

\subsubsection{Overall comparison}\label{subsec:estimation:overall}
\begin{figure}
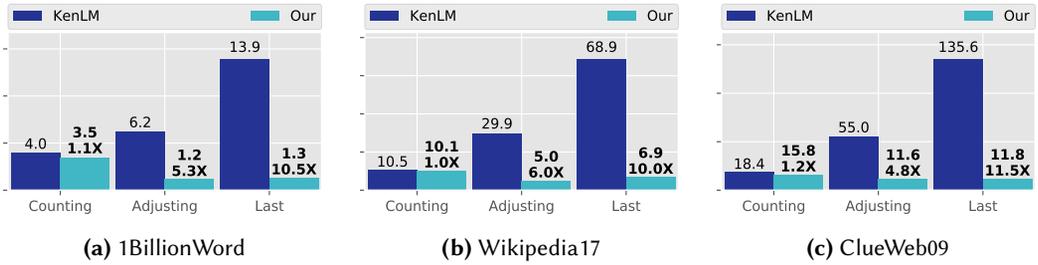

    \subfloat[$\BW$]{
    \includegraphics[scale=0.53]{{{plots/1Billion.overall_comparison}}}
    \label{fig:1Billion.overall_comparison}
    }
    \subfloat[$\WP$]{
    \includegraphics[scale=0.53]{{{plots/wikipedia.overall_comparison}}}
    \label{fig:wikipedia.overall_comparison}
    }
    \subfloat[$\CW$]{
    \includegraphics[scale=0.53]{{{plots/clueweb.overall_comparison}}}
    \label{fig:clueweb.overall_comparison}
    }   
    \caption{Time in minutes spent by $\ken$ and our algorithm at each step of estimation.}
    \label{fig:overall_comparison}
\end{figure}

In this final subsubsection we compare the performance of our solution, featuring all the optimizations that we have discussed before, against the state-of-the-art implementation of $\ts$ that is $\ken$.
The first comparison plots we show are illustrated in Figure~\ref{fig:overall_comparison}.
%As usual, upon each bar we report the number of minutes taken by the corresponding step. Below the minutes taken by our solution, we also indicate the speed-up achieved over $\ken$.
The plots strictly confirm the thesis of this paper. The round-trip performed by $\ts$, i.e., the sorting from suffix to context and then back from context to suffix (see Figure~\ref{fig:opt}), results in a severe penalty on the total running time of the estimation process: our improved $\os$ algorithm exploits the properties of the extracted $N$-gram strings in order to \emph{completely avoid} the round-trip.
Overall, this makes our approach run $4\times$, $4.9\times$ and $5.3\times$ faster than $\ken$, respectively on $\BW$, $\WP$ and $\CW$.
%In the following we discuss each step separately, also illustrating the software features adopted by $\ken$ in comparison with ours.
Let us now discuss each step separately.

\return
As already commented in Section~\ref{subsec:estimation:1-sort:counting}, the first step of Counting is performed similarly by the two algorithms and this is the reason why the corresponding running times are comparable.
In fact, both algorithms use a separate thread to sort the previously-formed block in parallel and flushing it to disk while input scanning takes place at the same time.
Both implementations also use open-addressing with linear probing.
The key difference lies in the fact that we sort in context order, whereas $\ken$ adopts suffix order. Another crucial difference is that our solution compresses the blocks to reduce the merging time in the next step.

During the Adjusting step, our approach computes the modified counts in context order on every output block formed during the merging process.
$\ken$ does the same but over suffix-sorted blocks, thus it has to write back to disk \emph{each $n$-gram, for $1 < n \leq N$}, along with its own modified count, in context order.
Since our approach re-computes the modified counts during the process of normalization itself, we only need to handle the $N$-grams and merge their blocks.
Instead, $\ken$ has to finally merge the blocks or all $n$-grams written to disk.
Although it exploits multiple threads (one for each order), the additional writes to disk and sorting operations cause $\ken$ be on average $5.3\times$ slower during this step than our approach.

During the Last step, normalization, interpolation and indexing are performed (Section~\ref{subsec:estimation:1-sort:normalization-interpolation} and~\ref{subsec:estimation:1-sort:joining-indexing}).
Again, we can observe an average speed-up of $10.6\times$.
Since our algorithm builds a compressed reverse trie index during the same step, we also sum to the time of $\ken$ the time it takes to build the same data structure, because the current implementation does not build the index during the same pass (although the possibility is advocated in the paper~\cite{HPCK13}).
To ensure fairness, the indexing time for $\ken$ is measured by excluding the time to write and parse the intermediate (ARPA) file on disk: it is anyway a significant amount of the total running time of $\ken$, equal to $7$, $31$ and $61$ minutes for, respectively, $\BW$, $\WP$ and $\CW$.
Apart from indexing, the rest of the time is spent in sorting again from context to suffix order, as needed for interpolation.
Both normalization and interpolation phases of $\ken$ exploits multi-threading, by using separate threads for each value of $n$.
In particular, two threads are used to compute the denominators and numerators of the quantities in Equation \ref{eq:norm} and \ref{eq:backoff}.
Again, recall that we only need to tackle $N$-grams because we consider the other $n$-gram strings implicitly, thus our implementation uses multiple threads for in-memory processing and a thread to asynchronously feed the CPU with input.

%total_speed_up: 4.01666666667
%total_speed_up: 4.97344461305
%total_speed_up: 5.3370212766

%\begin{figure}
%    \hspace*{-1.1cm}
%    \includegraphics[scale=0.7]{{{plots/wikipedia.IO_volume}}}
%    \caption{Gigabytes per second written on disk by $\ken$ and our algorithm when processing the $\WP$ dataset.}
%     \label{fig:IO_volume}
%\end{figure}

\begin{figure*}
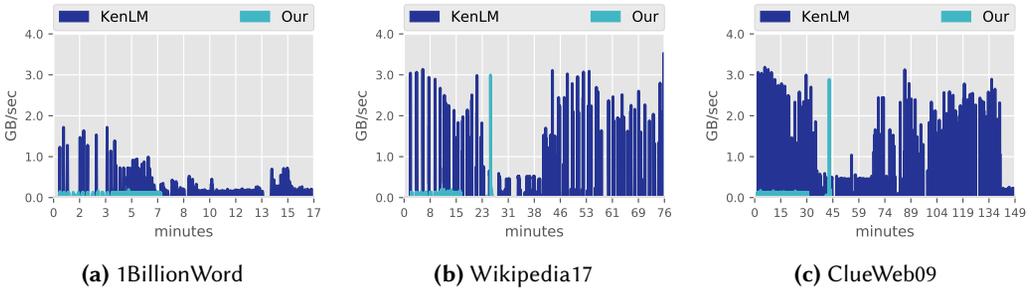

   \hspace*{-0.5cm}
   \subfloat[$\BW$]{
    \includegraphics[scale=0.55]{{{plots/1Billion.IO_volume}}}
   \label{fig:1Billion.IO_volume}
    }
   \subfloat[$\WP$]{
    \includegraphics[scale=0.55]{{{plots/wikipedia.IO_volume}}}
   \label{fig:wikipedia.IO_volume}
   }
   \subfloat[$\CW$]{
    \includegraphics[scale=0.55]{{{plots/clueweb.IO_volume}}}
   \label{fig:clueweb.IO_volume}
   }   
    \caption{Gigabytes per second written on disk by $\ken$ and our algorithm.}
    \label{fig:IO_volume}
\end{figure*}

\parag{Output volume}
Another way of visualizing the comparison between our solution and $\ken$ is to measure the number of bytes read/written per second from/to the disk by the two algorithms.
Figure~\ref{fig:IO_volume} shows the number of GB written per second on disk for each dataset.
We collect the statistic using the Linux utility \textsf{pidstat} with time interval of 1 second and matching the name of the executed task.
The volume for our construction also includes the one spent when flushing the compressed index to disk, whereas the volume for $\ken$ does not because the current implementation builds the index with a separate program.

The plots strictly match the results shown in Figure~\ref{fig:overall_comparison},
i.e., not surprisingly the improvement in running time is directly proportional to the quantity of data written to disk.
In fact, the area below the curve of our algorithm is
$\approx$6$\times$ less than the one of $\ken$ (20.4 GB vs. 124.5 GB) on $\BW$;
$\approx$5$\times$ less on $\WP$ (63.6 GB vs. 310.7 GB) and
$\approx$5.8$\times$ less on $\CW$ (88 GB vs. 514 GB).

%Compressed:
%1-sort area: 21446641.0;
%3-sort area: 130525350.0;
%num. points: 992;
%6X

%1-sort area: 66687992.0;
%3-sort area: 325769548.0;
%num. points: 4579;
%5X

%1-sort area: 92371338.0
%3-sort area: 539521156.0
%num. points: 8921
%5.8X

\section{Conclusions}\label{sec:conclusions}
In this paper we studied in depth the two problems lying at the core of language model applications, i.e., providing random access to $n$-gram probabilities and estimating such probabilities form large textual collections.
We focused on solving these two problems \emph{efficiently}.

Concerning the problem of indexing, we presented highly compact and fast indexes that achieve substantial performance improvements over the state-of-the-art approaches. In particular, our trie data structure exploits the succinctness of the Elias-Fano encoding by preserving the query processing speed of the fastest implementation in the literature.
We also introduced a context-based remapping technique for vocabulary tokens to further compress the trie data structure. On average, this technique improves compression by $28\%$ with a context of length $1$ and by $35\%$ with a context of length $2$, with only a slight penalty at query processing speed.

%Some interesting research problems naturally arise from the ideas described in the paper. We mention some of them. Is the frequency-based assignment of vocabulary tokens optimal for Elias-Fano?
%An interesting, yet hard, research problem could focus on devising an ID-assignment strategy with proven guarantees of optimality for Elias-Fano and for our context-based remapping technique.
%Additional efforts could also be spent on making trie searches even faster, e.g., by exploiting SIMD processor special instructions.

Concerning the problem of estimation, we described a novel algorithm that estimates unpruned, modified, Kneser-Ney language models in external memory.
Our approach sorts the extracted $N$-gram strings once, in \emph{context} order, and outputs the compressed trie data structure indexing the strings in \emph{suffix} order.
Our algorithms runs $4.5\times$ faster than the fastest state-of-the-art approach.
The improved performance of the algorithm derives from the exploitation of the properties of the extracted $N$-gram strings that relate context and suffix order and that are neglected by competitive approaches.

%Thanks to such properties, we showed that it is possible to operate over the context-sorted strings and, yet: (1) compute the Kneser-Ney modified counts in linear time and only taking space proportional to the vocabulary; (2) efficiently lay out the reverse trie data structure.

%Future research may improve parallelism and develop a distributed implementation of the described algorithm, targeting a cluster of multicore machines.

\begin{acks}

We thank Roberto Trani for his assistance with the textual collections of {\WP} and {\CW}.

This work was partially supported by the BIGDATAGRAPES project (grant agreement \#780751) that received funding from the European Union's Horizon 2020 research and innovation programme under the Information and Communication Technologies programme, and by the PEGASO project (POR FSE 2014-2020).

\end{acks}

\bibliographystyle{ACM-Reference-Format}
\bibliography{bibliography}

\end{document}